\documentclass[]{aastex631}

\usepackage{CJK}

    \shorttitle{Correcting Stellar Flare Frequency Distributions Detected by TESS and Kepler}
\shortauthors{Gao et al.}
\graphicspath{{./}{figures/}}
\begin{document}
\begin{CJK*}{UTF8}{gbsn}
\title{Correcting Stellar Flare Frequency Distributions Detected by TESS and Kepler}

\author[0000-0001-6643-2138]{Dong-Yang Gao(高东洋)}
\affiliation{School of Astronomy and Space Science, Nanjing University, Nanjing, 210023, China, \url{huigen@nju.edu.cn},\url{ming.yang@nju.edu.cn}}
\affiliation{Key Laboratory of Modern Astronomy and Astrophysics, Ministry of Education, Nanjing, 210023, China}
\affiliation{Shandong Key Laboratory of Optical Astronomy and Solar-Terrestrial Environment, Institute of Space Sciences, School of Space Science and Physics, Shandong University, Weihai, China}

\author[0000-0001-5162-1753]{Hui-Gen Liu(刘慧根)}
\author[0000-0002-6926-2872]{Ming Yang(杨明)}
\author[0000-0003-1680-2940]{Ji-Lin Zhou(周济林)}

\affiliation{School of Astronomy and Space Science, Nanjing University, Nanjing, 210023, China, \url{huigen@nju.edu.cn},\url{ming.yang@nju.edu.cn}}

\affiliation{Key Laboratory of Modern Astronomy and Astrophysics, Ministry of Education, Nanjing, 210023, China}

%\affiliation{School of Astronomy and Space Science, Nanjing University, Nanjing, 210023, China, \url{huigen@nju.edu.cn}}

%\affiliation{Key Laboratory of Modern Astronomy and Astrophysics, Ministry of Education, Nanjing, 210023, China}

%\affiliation{School of Astronomy and Space Science, Nanjing University, Nanjing, 210023, China, \url{huigen@nju.edu.cn}}
%\affiliation{Key Laboratory of Modern Astronomy and Astrophysics, Ministry of Education, Nanjing, 210023, China}

%\affiliation{School of Astronomy and Space Science, Nanjing University, Nanjing, 210023, China, \url{huigen@nju.edu.cn}}
%\affiliation{Key Laboratory of Modern Astronomy and Astrophysics, Ministry of Education, Nanjing, 210023, China}



\begin{abstract}

The habitability of planets is closely connected with the stellar activity, mainly the frequency of flares and the distribution of flare energy. Kepler and TESS find many flaring stars are detected via precise time-domain photometric data, and the frequency and energy distribution of stellar flares on different types of stars are studied statistically. However, the completeness and observational bias of detected flare events from different missions (e.g. Kepler and TESS) vary a lot. We use a unified data processing and detection method for flares events based on the light curve from Kepler and TESS. Then we perform injection and recovery tests in the original light curve of each star for each flare event to correct the completeness and energy of flares. Three samples of flaring stars are selected from Kepler and TESS, with rotating periods from 1 to $\sim$ 5 days. Adopting a hot-blackbody assumption, our results show that the cumulative flare frequency distributions (FFDs) of the same stars in Kepler and TESS bands tend to be 
consistent after correction, revealing a more natural flaring frequency and energy distribution. Our results also extend the low-energy limit in cumulative FFD fitting to $10^{31.5-33}$ erg on different types of stars. For solar-type stars, the average power-law index of cumulative FFD ($\alpha_{\rm cum}$)  is $-0.84$, which indicates that low-energy flares contribute less to the total flare energy. With a 
piecewise correlation between $\alpha_{\rm cum}$ and $T_{\rm eff}$, $\alpha_{\rm cum}$ first rises with $T_{\rm eff}$ from M2 to K1 stars, then slightly decreases for stars hotter than K1.

\end{abstract}



\section{Introduction} \label{sec:intro}

The light curves of Kepler \citep{2010Sci...327..977B} and TESS \citep{2014SPIE.9143E..20R} provide an outstanding chance to detect stellar variability, based on the high-precision and continuous photometry. Fruitful results have been achieved, e.g. exoplanets detection \footnote{\url{https://exoplanetarchive.ipac.caltech.edu}}; stellar flare detection that constrains the burst mechanism  (e.g. \citealp{2011AJ....141...50W,  2014ApJ...792...67C,2016ApJ...829...23D, 2019ApJS..241...29Y} etc.), and refined stellar parameters via stellar seismology (e.g. \citealp{2011A&A...534A...6C, 2012A&A...537A.134A, 2012A&A...543A..54A}, etc).

Stellar flares are closely related to the stellar activity. The explosion of the stellar flares can release enormous energy, mainly caused by the magnetic reconnection process in the coronal region (\citealp{1989SoPh..121..299P, 2010ARA&A..48..241B, 2016ASSL..427..373S}). Thus, detecting stellar flares give clues to understanding the stellar magnetic activity, which reflects the theory of stellar dynamo and internal structure. Stars of different spectral types have different convective depths. For example, low-mass stars (M$_*$ $<$ 0.3 M$_\sun$), i.e. stellar 
effective temperature cooler than M4, are fully convective,  while more massive stars ($0.3 M_\sun < M_{*} < 1.2 M_\sun$) have a radiation layer inside the convective layer. According to \citealp{2011AJ....141...50W} etc., flares on low-mass stars are usually  more frequent.    
%Thus, studying the stellar flares on different stars benefit understanding the internal structure of  

Stellar flares usually release intense ultraviolet radiation and are crucial to understanding the habitability of exoplanets \citep{2006Icar..183..491B}.  The far-ultraviolet radiation produced by flares can increase atmospheric erosion. Strong ultraviolet radiation reduces the habitability of rocky planets. Although Proxima Centauri b \citep{2016Natur.536..437A} is in the liquid-water habitable zone of its host star, the intensity of ultraviolet radiation on its surface is about 400 times that on the Earth, and there are many superflares ($\sim$ 8/year, $E_{\rm bol} \textgreater  10^{33}$ erg; \citealp{2004A&ARv..12...71G, 2007ApJS..173..673W, 2020MNRAS.494L..69A}). \citet{2018ApJ...860L..30H} even reported the first naked-eye superflare detected from Proxima Centauri. Thus, the habitability of Proxima Centauri b is still debatable \citep{2016A&A...596A.111R}. Research on ultraviolet energy and flare frequency is essential for studying habitability, especially for the rocky planets around M dwarfs.  However, a moderate near-ultraviolet flux may be necessary for prebiotic chemistry on the surface of rocky planets around low-mass stars \citep{2007Icar..192..582B}, e.g. TRAPPIST-1 planets \citep{2017Natur.542..456G} and Proxima Centauri b . %Proxima Centauri b is currently in the closest exoplanet system. 

To characterize the flare energy and frequency, the flare frequency distribution (FFD) is adopted and usually described by a power-law relation, i.e. $dN/dE \propto E^{\alpha}$ \citep{1985SoPh..100..465D}. The index $\alpha$ is utilized to constrain the magnetic activity for different stars. \citet{2019ApJS..241...29Y} presented a flare catalog of the Kepler mission using the long-cadence data in Data Release 25. They found that the FFDs from F-type stars to M-type stars obey this power-law relation with  $\alpha$  $\thicksim$ -2, indicating that they have a similar mechanism on generating flares. But the deviation of FFDs on A-type stars with $\alpha$  $\thicksim$ -1.1 implies a different mechanism.
% \citet{2020AJ....159...60G} identified and analyzed stellar flares with 2-minute cadence during the first two months of the TESS mission. They linked FFDs to criteria for prebiotic chemistry, atmospheric loss through coronal mass ejections, and ozone sterilization. 
 However, other works do not support similar $\alpha$ for stars of different type. The $\alpha$ of solar flares is about -1.75 (e.g., \citealp{1993SoPh..143..275C, 1995PASJ...47..251S, 2000ApJ...535.1047A}). For later-type main-sequence stars, measured $\alpha$ values range from -2.0 to -1.4 (e.g., \citealp{2018ApJ...858...55P}). The $\alpha$ value of -2.0 is critical. If $\alpha$ $\textless$ -2, the total flare energy distribution is dominated by low-energy flare events (e.g. \citealp{2003ApJ...582..423G}). In this case, the flares with low energy become important for redistributing the energy in the stellar atmosphere. For example, nanoflares with energies of $\sim 10^{24}$ erg have been suggested to heat the quiescent solar corona (\citealp{1988ApJ...330..474P, 1991SoPh..133..357H}). 

For the detection of flares, the current popular definition follows \citet{2015ApJ...814...35C} (e.g. \citealp{2016ApJ...829...23D, 2019ApJS..241...29Y}, etc.). To eliminate false-positive events, the flares require three consecutive points fulfilling the same criteria. Although the selected flares ensure the accuracy of the study of the morphological characteristics of a flare (e.g., \citealp{2014ApJ...797..122D, 2017ChA&A..41...32Y, 2021MNRAS.505L..79Y}), many real flare events with few sampling points are eliminated due to light curve cadence. Therefore, the number of low-energy or short-duration flares is underestimated. In addition, due to the detrending of a light curve to remove variable shape caused by the stellar rotation, and the uncertainty in reconstructing the flare shapes, the uncertainties in calculation of flare energy are $\sim 60\% - 65\%$ (e.g., \citealp{2019ApJS..241...29Y}). 

Since both the detection completeness of flare and the accuracy of energy calculation are important while fitting the FFDs. In this paper, we try to improve the completeness and the precision of energy calculation in detecting flares based on Kepler and TESS data, which can be used to correct the FFDs of different types of stars. Section \ref{sect:samples} describes the data and samples chosen in this work. Section \ref{sect:method}  describes the methods of data processing and FFD correction. Section \ref{sect:results} demonstrates our results 
based on our stellar samples. Finally, the summary and further discussion are presented in Section \ref{sect:summary}. 

\section{Data and Sample Selection} \label{sect:samples}
In this paper, we use the photometry data from Kepler and TESS to study stellar flares because of their high photometric precision and continuity. Both Kepler and TESS provide large samples of stellar light curves, which is the basis for studying the stellar flares statistically.

Kepler was launched in 2009 with the primary scientific goal of exploring the structure and diversity of planetary systems by surveying a large sample of stars in a region of the Cygnus and Lyra constellations of our Galaxy. 
It is a Schmidt telescope design with a 0.95 m aperture and a spectral bandpass from 400 nm to 900 nm \citep{2016ksci.rept....1V}.
The photometric precision was expected to be 20 ppm for a 12th magnitude G2V star with a 6.5 hr integration, though solar-type stars themselves turned out to be noisier than expected (\citealp{2011ApJS..197....6G, 2015AJ....150..133G}). 
It has monitored the targets continuously and simultaneously, with a sampling interval of up to 1 minute for short-cadence (SC) data and 29.4 minutes for long-cadence (LC) data. 
Only 5000 targets have SC data, and their average time of observation is about two months. More than 200,000 targets have LC data, covering the whole Kepler mission for 48 months (from Q1 to Q17).

TESS was launched in 2018 with the primary goal of searching for transiting Earth-sized planets around nearby and bright stars. Its four 10 cm optical cameras simultaneously observe an entire field of 24°× 96°, with a red-optical bandpass covering the wavelength range from about 600 to 1000 nm \citep{Tenenbaum2018TESSSD}. Note the TESS band is redder than the Kepler band. TESS measured light curves of over 200,000 preselected stars in its first two-year primary mission with a cadence of 2 minutes. Each sector of TESS covers about 27 days. 

In 2019 July, TESS moved its field of view toward the northern ecliptic hemisphere for its second-year mission and observed the sky monitored by Kepler.  
Therefore, it allows us to statistically analyze the same flaring stars from both Kepler and TESS \textbf{(e.g., \citealp{2020AJ....160...36D})}. 
Note that the photometric precision of different missions and the contamination of nearby stars caused by the large pixel scale of TESS (see Section \ref{subsect:method_findflares} and Section \ref{subsect:results_twomission}) mean that we need to deal with TESS data carefully.

\begin{figure}[ht!]
\plotone{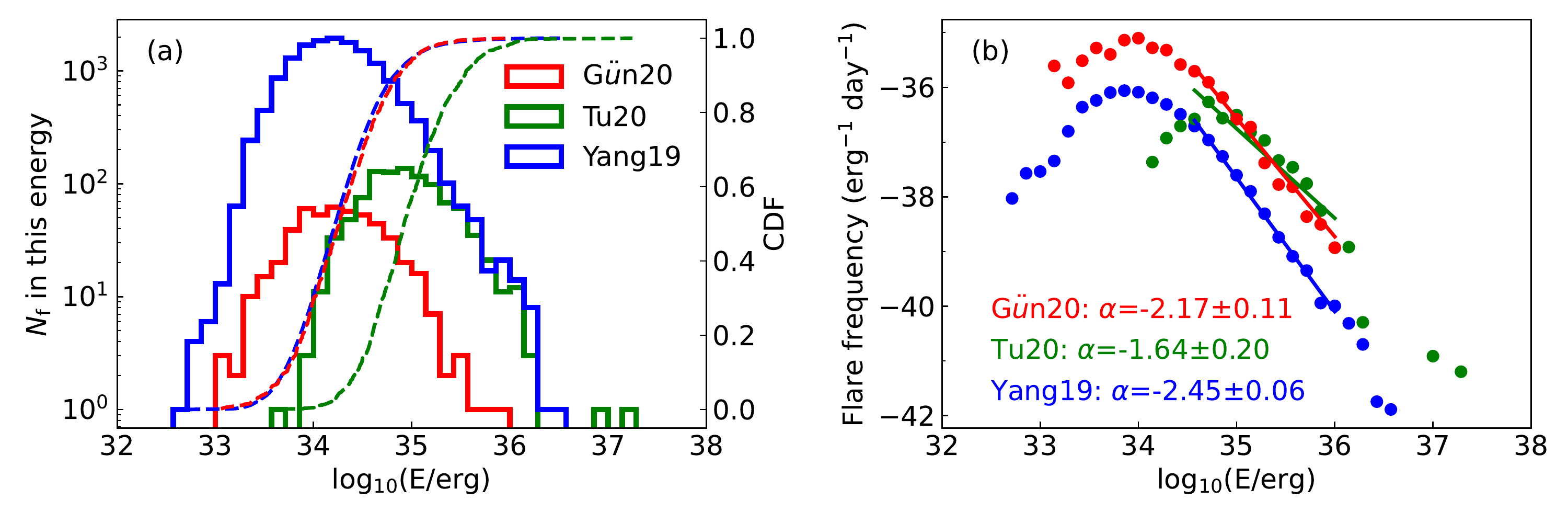}
\caption{Different flare catalogs of solar-type stars. (a) The histogram of 15,056 flares on 507 stars from Yang19 (solid blue), 989 superflares on 307 stars from Tu20 (solid green), and 502 flares on 107 stars from G$\ddot{u}$n20 (solid red). The blue, green, and red dashed lines are the cumulative distribution functions from Yang19, Tu20 and G$\ddot{u}$n20, respectively. N$_{\rm f}$ is the number of flares. (b) The blue, green, and red dots are the FFDs from Yang19, Tu20 and G$\ddot{u}$n20, respectively. The blue, green, and red solid lines are the fitted power laws of the respective FFDs in the energy range of $10^{34.5}-10^{36.0}$ erg. The FFDs of different catalogs using are fitted the equation \emph{$log_{10}(dN(E)/(dE\cdot dt))=\alpha \cdot log_{10}(E)+C$} (see Section \ref{subsect:method_fitcffd}). The text in the figure indicates the fitted parameter values and the errors of $\alpha$. %(c) Cumulative FFDs from different catalogs.
\label{fig:flarescatalog}}
\end{figure}

In previous works, \citet{2019ApJS..241...29Y} (hereafter \textbf{Yang19}) presented a flare catalog of the Kepler mission using the LC data of Data Release 25, comprising 3420 flaring stars and 162,262 flare events.
\citet{2020AJ....159...60G} (hereafter G$\ddot{u}$n20) performed a study of stellar flares for the 24,809 stars observed with a cadence of 2 minutes during the first two months of the TESS mission. 
\citet{2020ApJ...890...46T} (hereafter \textbf{Tu20}) studied superflares on solar-type stars from the first year's Observations from TESS. 
We collected the flare events on solar-type stars from their catalog and plotted the distribution of flare energy in Figure \ref{fig:flarescatalog}(a).
In a relatively complete detection range (flare energy between 10$^{34.5}$ and 10$^{36}$ erg), the power-law index ($\alpha$) of the FFDs from Yang19, Tu20 and G$\ddot{u}$n20 is -2.45$\pm$0.06, -1.64$\pm$0.20, and -2.17$\pm$0.11 respectively (see Figure \ref{fig:flarescatalog}(b)). 
Even in the same mission during the same epoch, different authors obtained different FFDs because of different values of detection completeness.

When using a complete method of detecting flare in light curves from different missions, the white-light FFDs of the same spectral type stars obtained from different missions should be similar and closer to the actual FFDs, especially in the flare energy region with better detection completeness for each mission. 
Therefore, we selected our samples from these three catalogs of flaring stars. 
First, we choose the samples observed by both Kepler and TESS to correct the difference due to bandpass, data cadence, photometric precision etc.. As can be seen in Section \ref{subsect:samples_keplerCMTLL}, most targets are M-type stars. Then we add stars of other spectral types to extend the sample to investigate the flare statistical characteristics of flares for different stars (see Section \ref{subsect:sample_diffstars}).
 
\subsection{Kepler Flaring Star with TESS Light Curve} \label{subsect:samples_keplerCMTLL}
Assuming that the actual FFD of a star is time-independent, we should obtain the same FFD on each star based on Kepler and TESS data. However, due to bandpass, photometric precision, and incompleteness of detection in different missions, the calculated number of flares and energy will lead to different FFDs via Kepler and TESS. We hope to find a method to correct the bias of FFDs in different missions. With the goal of correcting the FFDs to be closer to the actual values, we first choose the flaring stars observed by both Kepler and TESS missions.

We cross-matched \citet{2019ApJS..241...29Y}'s catalog with the TESS Target List\footnote{\url{https://tess.mit.edu/observations/target-lists/}} and got 193 targets within a distance of less than 1 arcsec. We sorted the target list according to the value of $N_{\rm f}$ (the number of flares in the Kepler mission) and selected stars with $N_{\rm flare}$ greater than 300 or stars with both $N_{\rm f}$ between 100 and 300 and Tmag between 12.13 and 14.13 which have relatively better precision in light curves of TESS. 
Finally, 13 stars are selected-- one G-type star, two K-type stars, and 10 M-type stars. The parameters of each star are listed in Table \ref{tab:stars}. These Kepler flaring stars with TESS light curves (hereafter denoted as Sample-1) have Tmag from 12.13 to 14.46 (see Table \ref{tab:stars}). These stars are faint for the TESS mission, and their photometric light curves from TESS are not as good as those from Kepler for detecting flares. For example, for the faintest star, TIC 27685142, a solar-type star with Tmag=14.46,  we can only detect 10 flare events using the TESS data (see Section \ref{subsect:method_findflares}). We cannot fit the cumulative FFD of the TESS mission for stars with few detected flares. Therefore, we verify the criterion while detecting flares in Section \ref{subsect:method_findflares}, and cut the sample more strictly in Section \ref{subsect:results_twomission} to exclude large uncertainties of the cumulative FFDs.
  
%\begin{longrotatetable}
\begin{deluxetable*}{cccccccccc}
%\tablenum{1}
\tablecaption{Stellar Parameters of Three Selected Samples. \label{tab:stars}}
\tablewidth{0pt}
\tablehead{ Samples &
\colhead{TIC ID} & \colhead{KIC ID} & \colhead{\textbf{GAIA ID}} & \colhead{Tmag} &
\colhead{Kmag} & \colhead{$T_{\rm eff}$ (K) $^{\rm a}$} & \colhead{\textbf{RUWE}} & \colhead{P$_{\rm rot}$ (days) $^{\rm b}$} 
& \colhead{$N_{\rm f}$ $^{\rm c}$} 
}
%\decimalcolnumbers
\startdata 
& 164670606 & 8935655 & 2106808834537729664 & 12.35 & 13.35  & 3371 & 1.19 & 2.00  & 491   \\
Sample-1  &27454242 & 12314646 & 2135580201979587072 & 13.15 & 14.14  & 3675   & 2.05 & 2.72 & 387 \\
 & ... & ... & ... & ... & ... & ... & ... & ...  \\
\hline
 & 102032397 & - & 4933439271456748032 & 10.59 & -  & 5435 & 1.07 & 3.70 & 4   \\
Sample-2 & 11046349 & - & 3251166758468483968& 9.42 & -  & 5410 & 1.00 & 3.33 & 2   \\
 & ... & ... & ... & ... & ... & ... & ... & ...   \\
\hline
& 150359500 & - & 5481843025342799104 & 9.75 & -  & 3286 & 1.10 & 1.03 & 109  \\
Sample-3 & - & 7131515 & 2077896248578080128 & - & 15.62  & 3902 & 1.10  & 3.86 & 401   \\
& ... & ... & ... & ... & ... & ... & ... & ...  \\
\enddata
\tablecomments{Sample-1: Kepler Flaring stars \citep{2019ApJS..241...29Y} with TESS light curves (13 stars, including six stars for rigorous analysis, see Section \ref{subsect:results_twomission}); Sample-2: solar-type TESS flaring stars from \citet{2020ApJ...890...46T} (84 stars); Sample-3: flaring stars with different spectral types from \citet{2020AJ....159...60G} (20 stars) and \citet{2019ApJS..241...29Y} (four stars). The flag ``-" indicates no value. \\
$^{\rm a}$ For stars in Sample-1 and Sample-2, the stellar parameters are adopted from {\tt\string astroquery.mast.Catalogs} ({\url{https://astroquery.readthedocs.io}}) from TESS input catalog (TIC v8) \citep{2022yCat.4039....0P}; For Kepler flaring stars in Sample-3, their stellar parameters are from \citet{2020AJ....159..280B}. \textbf{We also listed the stellar parameters from GAIA DR3 \citep{2022arXiv220605992F} in this table.}\\
$^{\rm b}$ Stellar rotation periods are taken from \citet{2020AJ....159...60G} and \citet{2019ApJS..241...29Y}, partly given by \citet{2013AA...560A...4R} and \citet{2014ApJS..211...24M}. \\
$^{\rm c}$ For Kepler flaring stars, ``$N_{\rm f}$" is taken from \citet{2019ApJS..241...29Y}; for TESS flaring stars, ``$N_{\rm f}$" is taken from \citet{2020AJ....159...60G} and \citet{2020ApJ...890...46T}; The entire table contains the number of flares we detected. \\
(This table is available in its entirety in machine-readable form.)}
\end{deluxetable*}
%\end{longrotatetable}

\subsection{Select of Stars for the Completeness of Spectral Type} \label{subsect:sample_diffstars}
In Section \ref{subsect:samples_keplerCMTLL}, we have selected 13 stars with different spectral types. As the most interesting case, the statistics of flare events for the solar-type stars are important for understanding the quiescent Sun, but Sample-1 only contains one solar-type star. Note we adopt stars with effective temperature between 5100 K and 6000 K as solar-type stars. To involve more solar-type stars, we select solar-type stars with a rotation period between 1 and 5 days and randomly choose 84 targets in \citet{2020ApJ...890...46T} (hereafter denoted as Sample-2). The cut of rotation period is because the burst frequency of flares is related to the star's rotation period (\citealp{2016ApJ...829...23D,2019ApJS..241...29Y,2020ApJ...890...46T}, etc.). Sample 2 will be used to investigate the flares characteristics of solar-type stars.

 Furthermore, to investigate the variation of a star’s FFD with spectral type, we need to add more stars with different spectral types. We select the stars from the catalogs in the papers of \citet{2020AJ....159...60G}, \citet{2020ApJ...890...46T} and \citet{2019ApJS..241...29Y} (see Table \ref{tab:stars}). The stars are classified into K4-K2, K8-K5, M2-M0, and M5-M3 according to their effective temperature via \citet{2013ApJS..208....9P}. For each stellar type, we select the top five stars with the highest flare frequency in \citet{2020AJ....159...60G}'s catalog, while the star with the highest flare frequency in \citet{2019ApJS..241...29Y}'s catalog is selected. In total, we select 24 stars as a sample including different spectral types (hereafter denoted as Sample-3).
 
In Sample-2 and Sample-3, the Tmag of TESS flaring stars varies from 7.78 to 12.14, while the Kmag of Kepler flaring stars in Sample-3 varies from 10.96 to 16.48. The light curves of these flaring stars are precise enough to detect flares (see Section \ref{subsect:method_findflares}).

To more accurately estimate flare energy based on stellar parameters, and to exclude contamination from binaries (e.g., \citealp{2019ApJ...876...58N}), we cross-matched these stars to GAIA DR3 \citep{2022arXiv220605992F}. We also listed the updated stellar parameters and renormalised unit weight error (RUWE) from GAIA DR3 in Table \ref{tab:stars} and applied them in Section \ref{sect:results}.

\section{Method}  \label{sect:method}
\begin{figure}[ht!]
\plotone{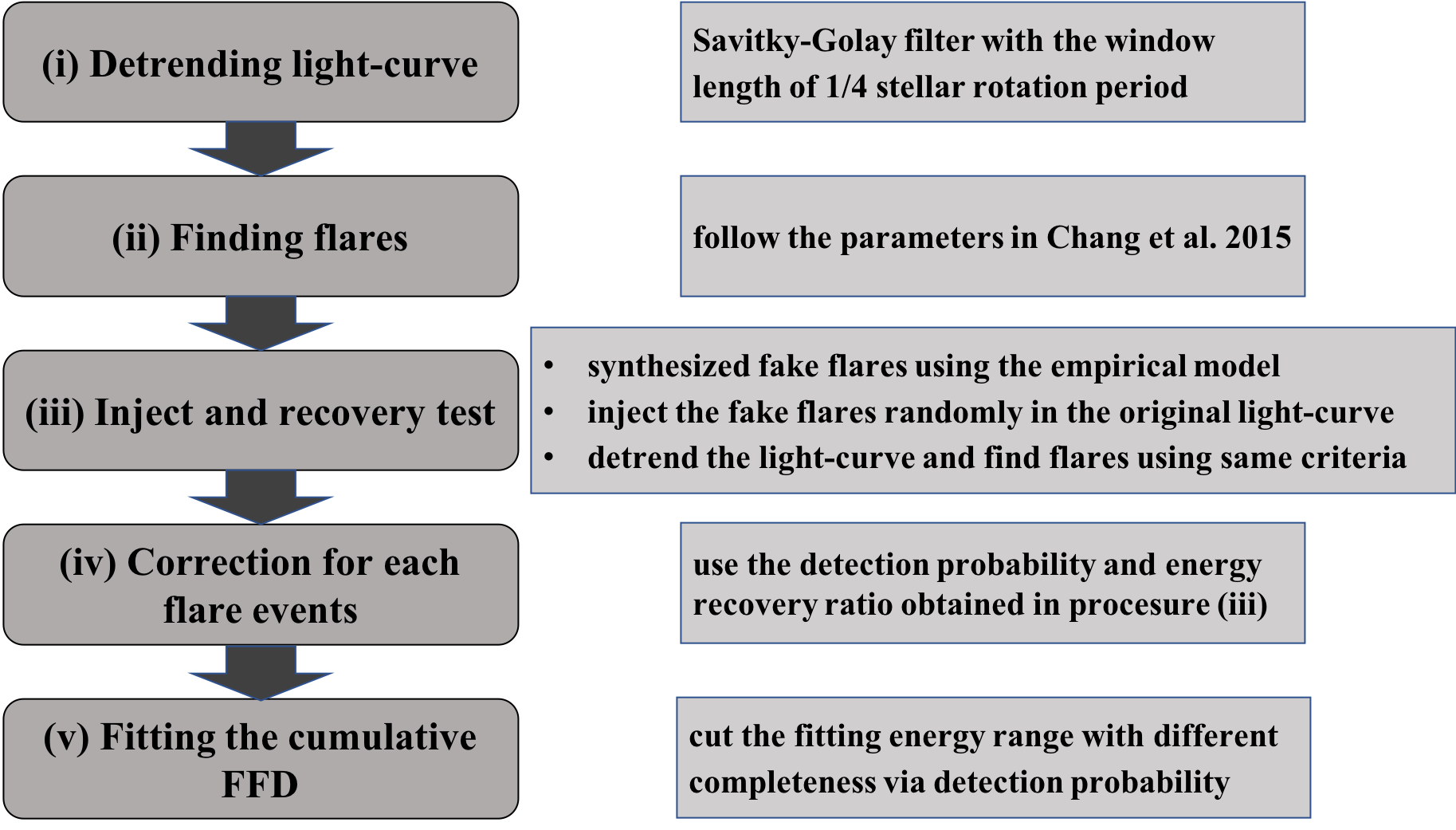}
\caption{The five procedures adopted in this paper. Note procedures (ii) and (iii) are based on the improved Python package {\tt\string AltaiPony} (\citealp{2016ApJ...829...23D, 2021A&A...645A..42I}).   \label{fig:procedure}}
\end{figure}

To correct the completeness of flare detection based on the data from Kepler and TESS, we reprocess the light curve uniformly and correct the detection efficiency by the following steps for each star: (i) detrend the light curve, (ii) find flare events, (iii) perform injection and recovery tests for each flare event, and (iv) correct the number and energy of flare events based on the detection probability and energy recovery ratio obtained from step (iii), respectively. Finally, (v) we fit the cumulative FFD to obtain the power-law parameters. Figure \ref{fig:procedure} demonstrates the procedures we adopted in this section, which will be introduced in detail one by one as follows. 

\subsection{Detrending the Light Curve} \label{subsect:method_detrendlc}
Flare detection and energy estimation are sensitive to the detrending method. In this subsection, we aim to correct the bias of the detrending methods in Kepler and TESS light curves, via a homogeneous detrending method.

As usual, we use the publicly available version of the Kepler or TESS photometry data, i.e. the pre-search data conditioning simple aperture photometry (PDCSAP) flux \citep{2012PASP..124.1000S}. The PDCSAP fluxes are obtained from a systematic reprocessing of the entire database using a Bayesian approach to remove systematic noises in both the short- and long-term light curves (\citealp{2014ApJ...797..121H}).

First, we smooth out the secular modulation signal. Note, the sampling stars are from the catalog of Kepler and TESS flaring stars (\citealp{2019ApJS..241...29Y, 2020ApJ...890...46T, 2020AJ....159...60G}), which has excluded out pulsating stars and binaries. Thus, the secular modulation signal is mainly caused by the spots rotating with star. 

We detrend the light curve using the Savitzky-Golay filter \citep{1964AnaCh..36.1627S} (see function {\tt\string  scipy.signal.savgol\_filter} function in SciPy package \citep{2020SciPy-NMeth}. The Savitzky-Golay filter is based on local polynomial least-squares fitting in the time series and can filter out noise while ensuring that the shape and width of the signal remain unchanged. 

There are two crucial parameters-- window length and polyorder-- when we use the Savitzky-Golay filter to detrend light curves. If the window length is too small, it will smooth out the characteristics of light variation of flares, especially long-lived ones. Conversely, if the window length is too large, it will not be able to smooth the rotation modulation signal for fast rotating stars. The optimal window length depends on the photometric error, sampling rate, light-curve trend, and filter order \citep{2018arXiv180810489S}. Thus, calculating the optimal window length requires a prior knowledge of the light-curve trend caused by stellar rotation modulation. 

\citet{2019ApJS..241...29Y} used the Lomb–Scargle periodogram (\citealp{1976Ap&SS..39..447L, 1982ApJ...263..835S}) to calculate the most significant period of each block, then smooth each block with the filter width of one-fourth of the most significant period. Their lower limit on filter width is 6 hr and the upper limit is 24 hr. 
We also set the window length as 1/4 of the stellar rotation period, while the stellar rotation period is from the previous catalog calculated through the Lomb–Scargle periodogram. Since all the stellar rotation periods in our samples are between 1 to 5 days, the  lower and upper limits on window length are 6 hr and 30 hr, respectively. We set the polynomial order as 3 to filter the PDCSAP light curve, as in the previous work (e.g. \citealp{2016ApJ...829...23D}).

The detrending method adopted in this paper differs from previous works of finding flaring stars. \citet{2020AJ....159...60G} computed the Lomb-Scargle periodogram and removed the periodic signal from light curve using a sine wave fit.
\citet{2020ApJ...890...46T} detected superflares without detrending the light curves, while a quadratic function is adopted to remove the long-term stellar variability to calculate the energy of superflares. %They did not detrend the light curve before finding superflares.

\subsection{Finding Flares and Energy Calculation} \label{subsect:method_findflares}
After detrending the light curve, we need to choose criteria to identify flare events and get the parameters of each flare. In this section, we will introduce our criteria for detecting flares for Kepler and TESS data. Furthermore, since the flare energy is crucial and is model-dependent, we introduce our method to calculate the flare energy based on the blackbody assumption.

\subsubsection{Flare Detection}
We identify flares based on the three parameters $N_{1,2,3}$  suggested by \citet{2015ApJ...814...35C} (the definitions of $N_{1,2,3}$ can be found in Appendix \ref{append:flaredfind}). \citet{2015ApJ...814...35C} recommend that $N_{1,2,3}$ should be at least larger than 3, 1, and 3, respectively. We focus on the completeness of flare detection in this work rather than a detailed study of flare morphology. Therefore, we do not verify false positives visually for detected flare events. 

%% The "ht!" tells LaTeX to put the figure "here" first, at the "top" next

% choose different N1 for lc with different precision;
To determine the parameters $N_{1,2,3}$, which are related to the photometric precision, we estimate the combined differential photometric precision (CDPP) using the Savitzky-Golay method (see the $\tt\string estimate\_cdpp$ function in the {\tt\string Lightkurve} package; \citep{2018ascl.soft12013L}). 
For these 13 Kepler flaring stars with TESS light curve with Tmag from 12.13 to 14.46, the CDPPs of the Kepler light curve are around 200 ppm, and then the $N_{1,2,3}$ values are set as 3, 2,  and 3, respectively. The CDPPs of the TESS mission for these targets are 1470-9480 ppm. To identify more flares with such photometric accuracy of TESS, the $N_{1,2}$ are set as 2, and $N_{3}$ is set as 2 to detect flares with shorter duration, when we detect flares from TESS light curves. \citet{2016ApJ...829...23D} also found that the flare recovery in Kepler is improved via long-cadence data and that adjusting the scale factor $N_{3}$ from 3 to 2 did not enhance the recovered flares obviously for short-cadence data. We also visually inspect the choice of parameter $N_{1,2,3}$, especially for the light curve of TESS, and believe that the flare events we found are confirmed.
%After correction using the detection probability and energy recovery ratio from injecting and recovery test (see Sect \ref{subsect:method_inject}) , the similar FFDs we obtained through TESS and Kepler mission also supported our parameter selection.

After we set the detection criteria of $N_{1,2,3}$, some flares can be detected at the beginning or end of each segment. These flares are suspected when such events are only partially observed or correlated with the detrending precision at the end or beginning of available light-curve data \citep{2015ApJ...814...35C}. Therefore, we discard flare events that overlapped with the 4 data points (0.005 days) and six data points (0.125 days), at both beginnings and ends of each light-curve segment, for the TESS and Kepler data, respectively. Additionally, there are impulsive outliers in the data marked as `512' (\citealp{Tenenbaum2018TESSSD}). We also abandon the flares that overlap with these outliers, which are usually doubted. Consequently, the available observation time is shortened in the FFD calculation. Take TESS data in Sector 1 as an instance: after excluding both the beginning and end of each segment in Sector 1, as well as the impulsive outliers, the available observational time is reduced from 27.88 to 25.36 days. We use the reduced observational time to calculate the FFD of each star.

\begin{figure}[ht!]
\plotone{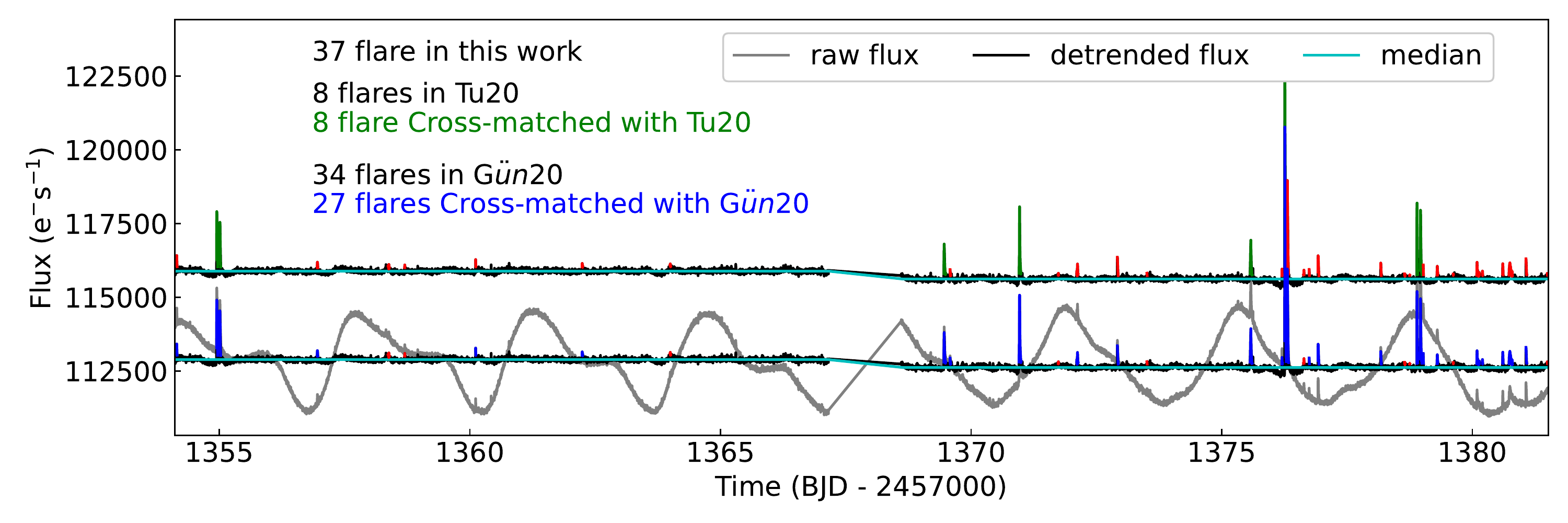}
\caption{The flares we found using our pipeline for TIC 206592394 in Sector 2 of the TESS mission. The red, green, and blue events show the flares we found. The green and the blue events are the same as in the catalogs of \citet{2020ApJ...890...46T} and \citet{2020AJ....159...60G} respectively. The numbers of flares we detected are greater than in previous works, and are annotated on the figure. \label{fig:flaresintwo}}
\end{figure}

We compare our flare events detected with those in previous works to validate the choice of $N_{1,2,3}$ adopted in this paper. We choose TIC 206592394 because both \citet{2020AJ....159...60G} and \citet{2020ApJ...890...46T} detect flares on this star.
The light curves of TIC 206592394 in Sector 2 are shown in Figure \ref{fig:flaresintwo}. The red, green, and blue events show the flares we found. The eight green and 27 blue flares are cross-matched with \citet{2020ApJ...890...46T}'s and \citet{2020AJ....159...60G}'s catalogs, respectively. The number of flares with low energy is slightly more than that from \citet{2020AJ....159...60G}'s catalog. Two things need to be addressed: (1) \citet{2020ApJ...890...46T} focus on superflares on solar-type stars, thus we recover their detection and detect 29 more flares; (2) for each outburst epoch, using a Gaussian process and Bayesian evidence, \citet{2020AJ....159...60G} sequentially add extra flares to ensure overlapping flares are distinguished even if the detection pipeline and visual inspection missed them. Although we do not distinguish the overlapped flares, $\sim$80$\%$ of flares in \citet{2020AJ....159...60G} are recovered by our pipeline for TIC 206592394. Thus, our pipeline has good completeness in detecting flare events compared with previous works.

\subsubsection{Flare Energy Calculation}
According to previous work, the bolometric energy of a flare can hard be calculate precisely, because of the complexity of the flare generation mechanism. The flares have different time-dependent spectra, even if flares on the same star. The flare energies may depend on the area and magnetic structure of starspots (\citealp{2021ApJ...906...72O}) during different flare events. At present, a common method for calculating the flare energy is to assume that the hot-blackbody radiation with a constant temperature of 9000K or 10,000K in the active area of the stellar surface, and that only the active area changes during the flares. The assumption is not correct, since the temperature of the stellar flares will change significantly during the injection and release of energy. Furthermore, the hot-blackbody assumption cannot well represent the Balmer jump and underestimates the  energy radiated in the near-UV region by factors of 2-3 \citep{2016ApJ...820...95K,2018csss.confE..42K,2019ApJ...871..167K}. Accurate estimation of flare energy requires more observations of multiple wave bands (e.g., \citealp{2016ApJ...820...89F,2022ApJ...929..169R}) and observation of time series for each event.

For comparison with previous work, we simply adopt the hot-blackbody assumption in this paper. The FFDs from F to M stars all obey a power-law relation with $\alpha$ $\sim$ 2, indicating that they have the similar burst mechanism of magnetic reconnection (e.g., \citealp{2019ApJS..241...29Y,2020AJ....159...60G,2021MNRAS.504.3246J}). Then, we assume that the same spectral shape among flares on different stars with different flare energies, i.e. temperatures of flares are all set as 9000 K (\citealp{2013ApJS..209....5S,2016ApJ...829...23D,2018ApJ...859...87Y,2019ApJS..241...29Y,2020AJ....159...60G}). 
There is another method to calculate the energy of flares, e.g. \citet{2015ApJ...798...92W} and \citet{2020ApJ...890...46T} integrate the energy based on the enhancement in the light curve, assuming the enhancement is constant in all wavelengths. Refer to Appendix \ref{append:energycalc} for exact energy calculation formulae and differences between the two methods.

After finding flare events, the pipeline records the amplitude (\emph{$Amp.$}), equivalent duration (\emph{$ED$}) and duration (\emph{$Dur.$}) of flares (\citealp{2016ApJ...829...23D,2021A&A...645A..42I}).
These parameters can be used to calculate the flare energy in the observational band. 
Since the energy calculation depends on the parameters $ED$, the uncertainty is sensitive to photometric precision, i.e. the recorded error of $ED$ is from the uncertainty on the flare flux values (see \citealp{2016ApJ...829...23D} Eqn. (2) for details).  For example, TIC 312590891 in Sample-2 with Tmag=9.45 (CDPP=255 ppm) has a median error $\delta_{\rm ED}=0.076$ s for all its flares found in TESS data, while a fainter star, TIC 27454242 in Sample-1 with Tmag=13.15 (CDPP=2554 ppm), has a larger median error $\delta_{\rm ED}=0.264$ s due to the 10 times larger photometric uncertainty. Additionally, TIC 27454242 is also observed by Kepler (KIC 12314646) with a much better photometric precision of 159 ppm. Thus, the $\delta_{\rm ED}$ from Kepler data is reduced to 0.049 seconds.  Obviously, for the flare events with the same $ED$, the worse photometric precision causes larger $\delta_{\rm ED}$, or the larger energy uncertainty. 

 Besides the photometric error, the uncertainties of the stellar parameters (e.g. $T_{\rm eff}$, $R_*$) also contribute to the energy uncertainty. Thus the calculated energy may be not precise for individual flare events. However, we focus on the statistical results of FFDs in this paper, especially the slope of FFDs, which is not sensitive to a consistent deviation due to uncertainties of stellar properties. Therefore we do not consider the uncertainty of calculated flare energy .

\subsubsection{Validation of Flare Detection and Energy Calculation via Our Method} \label{subsubsect:validation_method}
\begin{figure}[ht!]
\centering
\plotone{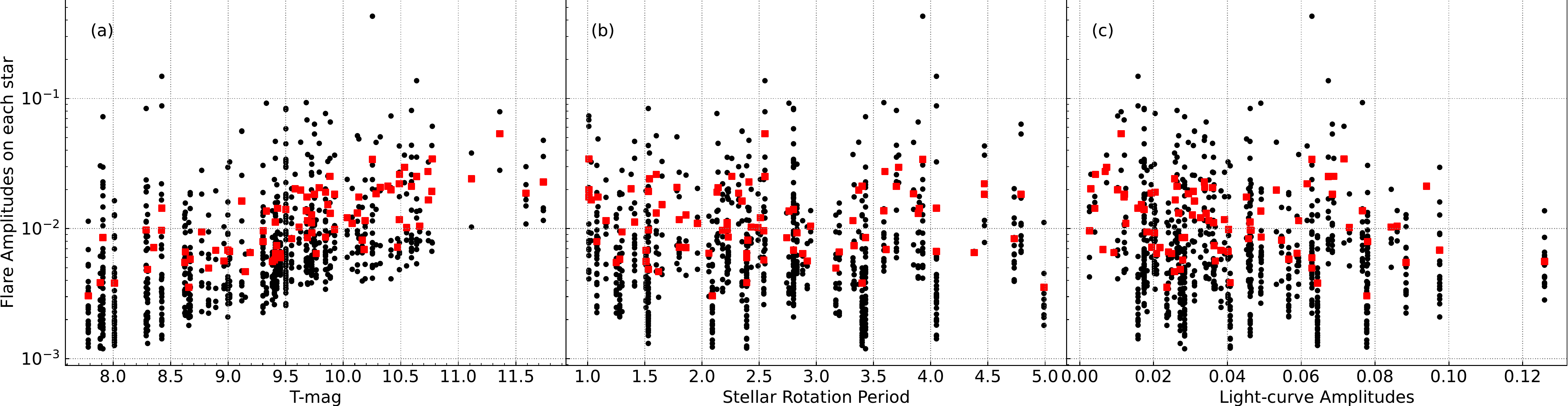}
\caption{Scatter plot of the normalized flare amplitudes  vs. different stellar parameters in Sample-2. The black dots mark each flare and the red squares mark the average values on each star.  
\label{fig:flareVSpars}} 
\end{figure}

To verify the completeness of flare detection, especially for fast rotating stars with various periodic light-curve amplitudes (e.g., \citealp{2021ApJ...906...72O}), we analyze the distribution of flare parameters with stellar parameters. The detrending light curve procedure is related to the stellar rotation period and the periodic amplitude of the light curve (LCAmp).  The Tmag of the star indicates the photometric accuracy, which is directly related to the procedure for finding flares. Stellar rotation periods are taken from \citet{2020AJ....159...60G} for TESS flaring stars, and from \citet{2013AA...560A...4R} and \citet{2014ApJS..211...24M} for Kepler flaring stars. We obtain the $LCAmp$ following \citet{2013MNRAS.432.1203M}.  Panel (a-c) of Figure. \ref{fig:flareVSpars} shows the distribution of the normalized flare amplitude with the stellar parameters (Tmag, $P_{\rm rot}$, $LCAmp$) in Sample-2.  Obviously, the normalized amplitudes of detected flares are positively correlated with the T-mag, but they have no obvious correlation with $P_{\rm rot}$ and $LCAmp$. The results indicate that the detrending method remove the variation module in light curves well.

In order to validate the flare detection and energy calculation via our method, we compare the detected flare events in Sample-2 and Sample-3 with previous work. For this comparison, we use the same stellar parameters as in the previous work. In Section \ref{sect:results}, we used the latest stellar parameters from the GAIA DR3 catalog.

\begin{figure}[ht!]
\plotone{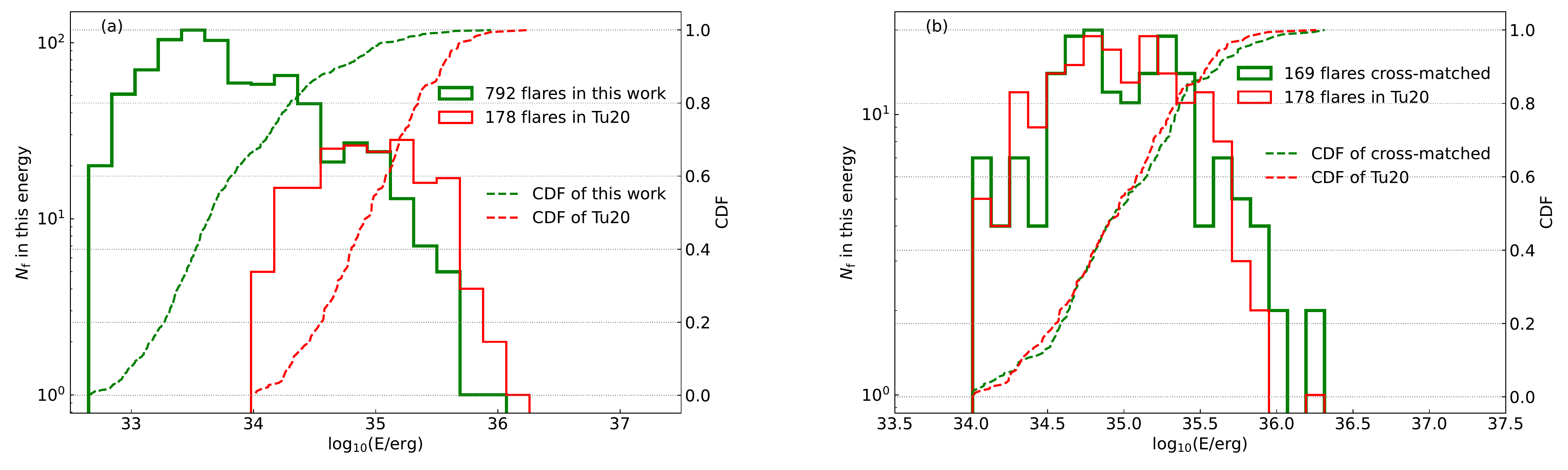}
\caption{(a) The histogram of 792 flare events we detected in Sample-2 (solid green) and 178 superflares in \citet{2020ApJ...890...46T} (solid red). Note that the energy calculation is based on the blackbody assumption and is not corrected in panel (a). The dashed green and red lines are the CDFs of all flares we detected and those from \citet{2020ApJ...890...46T}, respectively. (b) Similar to panel (a), the histogram of 178 superflares from \citet{2020ApJ...890...46T} (solid red) and the histogram of 169 cross-matched flare events we detected in Sample-2 (solid green). Note the energy correction is added in panel (b) to compare with Tu20's results  (see the text and Appendix \ref{append:energycalc} for detail). 
\label{fig:crossmatchTu}}
\end{figure}

We detect 792 flare events on Sample-2 stars, including $\sim$95\% of superflares in the catalog of \citet{2020ApJ...890...46T}. i.e. 169 flare events are cross-matched with the previous catalog of superflares, which includes 178 events. 
Figure \ref{fig:crossmatchTu}(a) shows the histogram and cumulative distribution function (CDF) of flares in \citet{2020ApJ...890...46T} and all flares we detected. More flares with energy below 10$^{34}$ erg are detected in our methods. 

Note that we calculate the energy of flare using the assumption of the hot-blackbody radiation, while \citet{2020ApJ...890...46T} integrate the energy based on the enhancement in the light curve (see Appendix \ref{append:energycalc} for detail about the two ways to calculate flare energy). In order to compare the similarity of flares between this work and \citet{2020ApJ...890...46T}, we correct the energy of flare events we detected via a factor $r$ in Eqn. \ref{equation:ratiooftwo} to correct the difference due to the two calculation methods. Figure \ref{fig:crossmatchTu}(b) shows the histogram in \citet{2020ApJ...890...46T} (green line) and the cross-matched flares we detected (red line). We also 
use a Kolmogorov–Smirnov (K-S) test to test the similarity of the two distributions, and this returns a P-value of 0.77, proving that our detection of superflares has nearly the same completeness and energy distribution as \citet{2020ApJ...890...46T}.

\begin{figure}[ht!]
\centering
\plotone{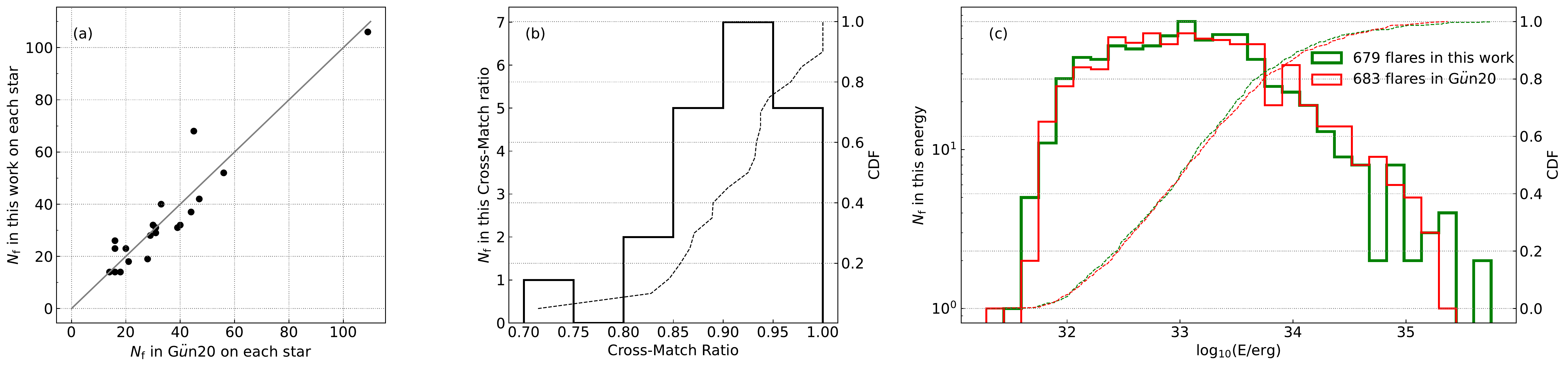}
\caption{Comparison of the flare events and energy distribution with \citet{2020AJ....159...60G}'s catalog for 20 TESS flaring stars in Sample-3. (a) The number of flares we detected vs. the number of flares from \citet{2020AJ....159...60G} for each star. (b) The histogram distribution (solid black line) and the CDF (dashed black line) of the match ratio between the flares we detected and flares from \citet{2020AJ....159...60G}. The ratio 
for each star is the fraction of flare events we detected in \citet{2020AJ....159...60G}’s catalog. (c) The histogram of flares in \citet{2020AJ....159...60G} (solid red line) and of those we detected (solid green line) after  correction (see text for details). The dashed green and red lines are CDFs of all flares. Note that \citet{2020AJ....159...60G} distinguish some flares with long duration as several overlapping flare events with lower energies, which we do not consider here. 
\label{fig:crossratio}} 
\end{figure}

In Sample-3, there are 20 TESS flaring stars from \citet{2020AJ....159...60G} and four Kepler flaring stars from \citet{2019ApJS..241...29Y}. We focus on the 20 TESS flaring stars to make the comparison.
For each TESS flaring star in Sample-3, we cross-matched the flare events detected by us and \citet{2020AJ....159...60G}. 
Figure \ref{fig:crossratio}(a) shows the number of flares we detected is consistent with the number of flares from \citet{2020AJ....159...60G} for each star. 
For all 20 TESS flare stars, we detect more than 70$\%$ of their flare events in \citet{2020AJ....159...60G}'s catalog, while even in 12 stars we detect more than 90$\%$ of flare events (Figure \ref{fig:crossratio}(b)).

To confirm the reliability of our flare detection and energy calculation, we also compare the flare distribution as a function of energy detected by our pipeline with that detected by \citet{2020AJ....159...60G}, as shown in Figure \ref{fig:crossratio}(c).
There are 683 events in \citet{2020AJ....159...60G}'s catalog, while we detected 679 events. 624 (92$\%$) of flares we detected are cross-matched with \citet{2020AJ....159...60G}'s catalog successfully. Here, we want to recall that \citet{2020AJ....159...60G} distinguish some flares with long duration as several overlapping flare events with lower energies, which we do not consider here. 

Note that the parameter settings we used, while integrating the flare light curves, are different from those used by \citet{2020AJ....159...60G} (see {\tt\string allesfitter} package (\citealp{2019zndo...3552952G}) and \citet{2020AJ....159...60G}'s paper for details). Thus the flare energies we obtained are roughly half of \citet{2020AJ....159...60G}'s results, although the consistent deviation in ratio is not crucial for the slope of FFDs. To compare the flare energy distribution with \citet{2020AJ....159...60G}, we add a correction factor of 2 in Figure \ref{fig:crossratio}(c). The K-S test between flares from \citet{2020AJ....159...60G} and ours returns a P-value of 0.63, which shows that our detection completeness and energy calculation are consistent with \citet{2020AJ....159...60G}. 

We conclude that the flare events we detected with higher energy are relatively consistent with previous works in both Sample-2 and Sample-3, which validates the detection method adopted by this work. Furthermore, we detect more flares with lower energies and can improve the completeness of flares to constrain the slope of the distribution with energy ($\alpha$, see Section \ref{subsect:method_fitcffd}).

\subsection{Inject and recovery test} \label{subsect:method_inject}
As seen in section \ref{subsect:method_findflares}, flares with higher energy are detected more completely in different works. However, flares with lower energies are not completely detected. Therefore, we perform injection and recovery tests to correct the completeness via the detection probability and improve the energy estimation via the energy recovery ratio.  

We synthesize flare shape with exactly the same amplitude and duration of each flare we detected in all samples, using the empirical model of \citet{2014ApJ...797..122D}. Then, we inject synthesized flare events at random epochs in the original light curve of the corresponding star. To avoid the overlapping of flare events, we limit the injection frequency to 0.5 per day, and make sure each segment has at least one injection. The flares are set individually and do not overlap with each other.
% ??data sampling  details??; 
We generate $\sim$2000 injections for each flare from TESS, and $\sim$4000 to 5000 injections for each Kepler flare. 
Then, we detrend the light curve with injected flares and find flares using the methods introduced in Section \ref{subsect:method_detrendlc} and Section \ref{subsect:method_findflares}.
Based on the  simulation of injection and recovery, the detection probability of each flare is defined as the fraction of recovered flare events, while the energy recovery ratio of each flare is the average of the ratio of recovered energy to the energy we injected initially. Finally, we obtain the detection probability and energy recovery ratio for each flare event based on a statistical analysis of the synthetic injection and recovery data (see the  {\tt\string characterize\_flares}, using  {\tt\string AltaiPony} package (\citealp{2016ApJ...829...23D,2021A&A...645A..42I}).

\begin{figure}[ht!]
\plotone{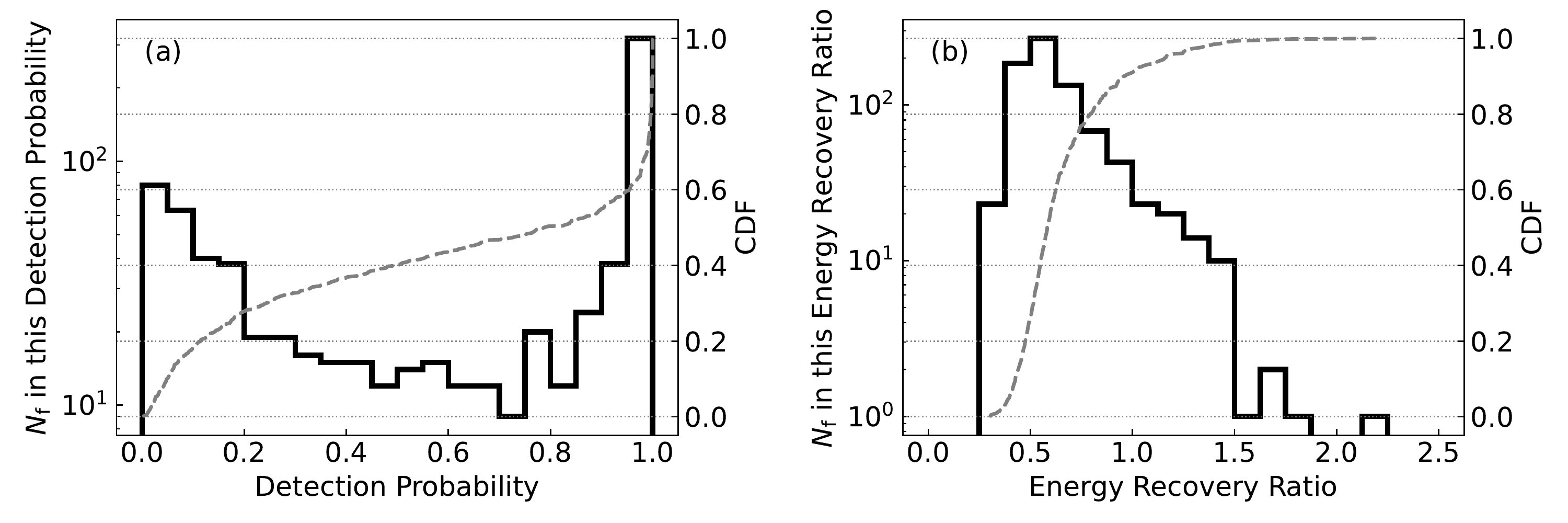}
\caption{Detection probability and energy recovery ratio of a total of 792 flare events on the 84 TESS flaring stars in Sample-2. (a) The histogram (solid black line) and CDF (dashed gray line) of the detection probability of these flare events. (b) The histogram (solid black line) and  CDF (dashed gray line) of the energy recovery ratio of these flare events. 
\label{fig:histofprob}}
\end{figure}

We analyze the detection probability and energy recovery ratio of 84 targets in \citet{2020ApJ...890...46T} (Sample-2). 
The total 792 flare events we detected on these 84 TESS flaring stars returned a detection probability and energy recovery ratio for our injection and recovery test.  Figure \ref{fig:histofprob} (a) and (b) show the histogram and CDF of detection probability and energy recovery ratio. 
50$\%$ of flare events have a detection probability of 0.8 or more, while 30$\%$ of flares have a detection probability of less than 0.2. Thus the number of flares detected with low detection probability is incomplete and needs to be corrected to estimate the real frequency.  
Furthermore, 60$\%$ of flare events have an energy recovery ratio less than 0.625, i.e. their energies are underestimated by at least 37.5$\%$. Note also that 9$\%$ of flare events that have an energy recovery ratio of more than 1.0, which means their total energies are overestimated. Therefore, the recovered energy probably deviates from the real value probably and needs to be corrected according to the energy recovery ratio.

Usually, lower-energy flares lead to lower detection probabilities and larger energy uncertainties due to lower signal-to-noise ratio. As seen in Figure \ref{fig:fitrange} (a) and (b), detection probabilities increase with the flare energy, and energy recovery ratios decrease with the flare energy in the case of a Kepler flaring star KIC 7131515. 

\subsection{Correction for Each Flare Event} \label{subsect:method_correction}
% general intro. of details to correct the flare events;
We denote the detection probability of a flare event as $P_{\rm det}$ and the energy recovery ratio as $R_{\rm rec}$. Then the actual number of flare events should be enlarged by a factor of $1/P_{\rm det}$. Assuming a flare event with an energy of $E_{\rm recovery}$, the actual energy should be corrected as $E_{\rm correct} = E_{\rm recovery}/R_{\rm rec}$. These corrections of numbers of flares and energies are adopted hereafter.

Note that if the detection probability or the energy recovery ratio is too small, the correction factor will return a large and unreliable value. Thus to avoid these extreme cases, we set a lower limit of 0.01. I.e. if the calculated probability or recovery ratio of a flare is less than 0.01, we skip over the correction to keep its energy and number as the original values and leave them as incomplete flares. However, only five flares (0.6\%) in Sample-2 have such small $P_{\rm det}$ and the minimum $R_{\rm rec}$ is 0.3. 

Finally, we make corrections according to each flare event's recovery probability and energy recovery ratio, flare by flare, star by star. 

\subsection{Fitting the flare frequency distribution} \label{subsect:method_fitcffd}
Flares with a power-law distribution in energy (\citealp{1976ApJS...30...85L}) are described by:
\begin{equation}
dN(E) = kE^\alpha dE dt \label{equation:ffd}, 
\end{equation}
where $N$ is the number of flares that occur in a given observational period $dt$, $E$ is the flare energy, $dt$ is the time spanning of observation, $k$ is a constant of proportionality, and $\alpha$ is the power-law index.

One can obtain the standard cumulative FFD by integrating the number of flares in unit time with energy range greater than a certain value. i.e. the frequency of flares (denoted by $\nu$) with energy $\geq E$ per day should be
\begin{equation}
log_{10}(\nu) = \beta + \alpha_{\rm cum}log_{10}(E) \label{equation:cffd},
\end{equation}
where \emph{$\beta =log_{10}(\frac{k}{1+\alpha})$} and $\alpha_{\rm cum}$=$\alpha$+1 ( \citealp{2014ApJ...797..121H, 2021MNRAS.504.3246J}). The power-law index  $\alpha_{\rm cum}$ represents the slope of the cumulative FFD. If a star has $\alpha_{\rm cum} \textless$ -1 ( i.e. $\alpha  \textless$ -2), the low-energy flares contribute the majority of total energy emitted by flares. Conversely, on the stars with $\alpha_{\rm cum}$ $\textgreater$ -1 (i.e. $\alpha \textgreater$ -2), the high-energy flares dominate the total energy (\citealp{2018ApJ...858...55P,2021MNRAS.504.3246J}).  

Note that the detected FFD is not exactly a power-low distribution, and the fitted power-law indices do not sarisfy the equation of $\alpha_{\rm cum}$=$\alpha$+1 \citep{2009MNRAS.395..931M}. Thus, we cannot directly transfer $\alpha_{\rm cum}$ to $\alpha$. In Section \ref{sect:results}, we will fit either $\alpha_{\rm cum}$ or $\alpha$ consistently with previous works for comparison. 

\begin{figure}[ht!]
%\begin{figure}[htb]
\centering
\includegraphics[width=0.75\linewidth]{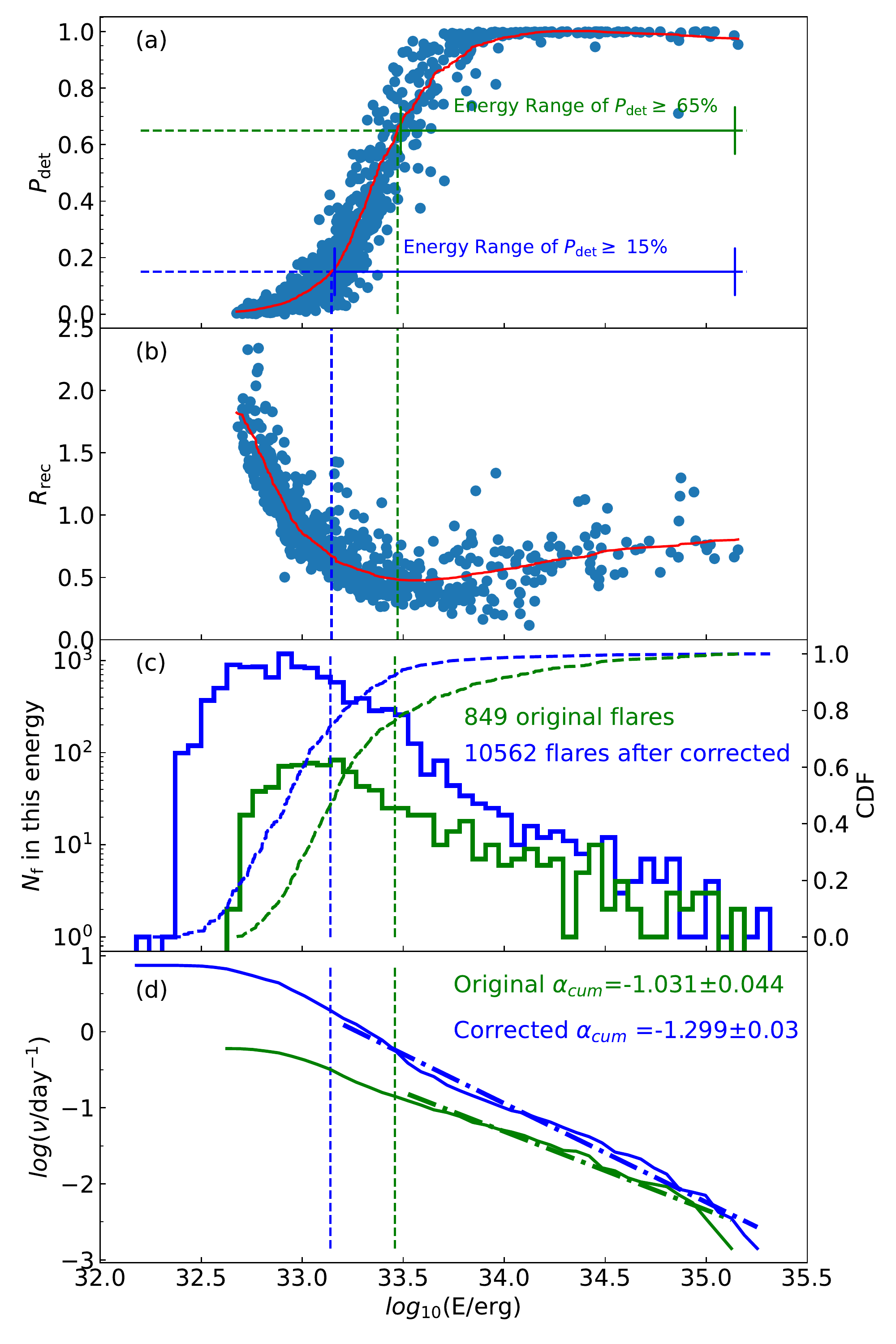}
%\plotone{fitrange.png}
\caption{The example of KIC 7131515. (a) The detected probability varying with flare energy. The red line is the smoothed curve of the scatter; the green and blue dashed horizontal lines indicate the value of $P_{\rm det}=65\%$ and 15\%, respectively. We choose the lower energy limit for fitting the original and corrected cumulative FFDs, based on the two critical values of $P_{\rm det}$ (the green and blue vertical lines, the same for other panels). (b) The energy recovery ratio varying with flare energy. (c) The histogram of flares before (solid green line) and after (solid blue line) correction. The green and blue dashed lines are the CDF of flares before and after correction, respectively. (d) Cumulative FFDs before (solid green line) and after (solid blue line) correction. The dashed-dotted lines shows the relationship fitted with Eqn. \ref{equation:cffd}. 
\label{fig:fitrange}}
\end{figure}

Since the fitted power-law index depends on the energy range we choose, the choice of energy range should be complete for flare detection. As described in Section \ref{subsect:method_correction}, we correct the completeness of flares via the detection probability $P_{\rm det}$. For a given flare star, we can plot $P_{\rm det}$ of each flare (e.g. Figure \ref{fig:fitrange} (a)) and obtain a smoothed correlation curve via a Savitzky-Golay filter. We set two lower limits of $P_{\rm det}$, i.e. 0.65 and 0.15, to preserve the flare samples with high completeness and with moderate completeness. Then the number and energy of flares in three samples are corrected to fit the FFD. 

We take KIC 7131515 as an instance. As shown in Figure \ref{fig:fitrange} (c), the number of flares we detected after correction is obviously more than that before correction, and the energy extends as low as $10^{32.5}$ erg. For flares with energy $\ge 10^{34} erg$, the histograms of corrected and original flares are similar, while those with energy 10$^{32.5-34.0}$ erg are very different. 

The power-law index of cumulative FFDs of each star before and after correction is listed in Table \ref{tab:alps_s1} (Sample-1) and Table \ref{tab:alps_s23} (Sample-2 and Sample-3).  The electronic version of Table 1-3 can be found on the website\footnote{\url{https://github.com/GaoDongYang2022/Correcting-Stellar-FFDs-Detected-by-TESS-and-Kepler}} . We will demonstrate the results in the next section.

%\begin{longrotatetable}
\begin{deluxetable*}{C|CC|CC|CC|CC}
\tablecaption{The power-law parameters for fitting cumulative FFDs before and after correction for flaring stars in Sample-1.\label{tab:alps_s1}}
%\tablewidth{700pt}
\tabletypesize{\scriptsize}
\tablehead{
\multicolumn{1}{c}{Star} & 
\multicolumn{2}{c}{TESS Original RangeAll} $^a$ &
\multicolumn{2}{c}{TESS Corrected RangeAll} $^a$ & 
\multicolumn{2}{c}{Kepler Original Range65}  &
\multicolumn{2}{c}{Kepler Corrected Range15}   \\
\colhead{TICID}  & \colhead{$\alpha_{\rm cum}$}  & \colhead{$\beta$} &\colhead{$\alpha_{\rm cum}$}  & \colhead{$\beta$}  & \colhead{$\alpha_{\rm cum}$}  & \colhead{$\beta$}  & \colhead{$\alpha_{\rm cum}$}  & \colhead{$\beta$} 
} 
\startdata
164670606  & -0.848\pm0.074 & 27.19\pm2.38  & -0.829\pm0.114 & 26.87\pm3.68  & -1.331\pm0.035 & 43.02\pm1.17  & -1.585\pm0.037 & 51.84\pm1.25  \\
27454242  & -0.514\pm0.037 & 16.73\pm1.25  & -1.385\pm0.272 & 47.00\pm9.19  & -1.326\pm0.027 & 44.29\pm0.93  & -1.287\pm0.017 & 43.27\pm0.58  \\
 ...  & ... & ... & ... & ... & ... & ... & ... & ... \\
\enddata
\tablecomments{``RangeAll'' means the fitting within the whole energy range; ``Range65'' means the fitting within the energy range of $P_{\rm det} \geq65\%$; ``Range15'' means the fitting within the energy range of $P_{\rm det} \geq15\%$;\\
$^{\rm a}$ Because of the requirement of sufficient energy bins, we only use the whole energy range to fit the cumulative FFDs obtained from the TESS data for Sample-1 stars. The fitting of FFD fails for ``Range65'' and  ``Range15'' due to the limited number of energy bins.
For faint stars in Sample-1, considering their worse photometric precision from TESS, we correct the flares with $P_{\rm det} \geq 0.001$ to enlarge the numbers of flares. (see Section \ref{subsect:method_correction})\\
(This table is available in its entirety in machine-readable form.)}
\end{deluxetable*}
%\end{longrotatetable}

%\begin{longrotatetable}
\begin{deluxetable*}{ccc|cc|cc|cc}
\tablecaption{The power-law parameters for fitting cumulative FFDs before and after correction for solar-type stars (Sample-2) and for stars of different types (Sample-3).\label{tab:alps_s23}}
%\tablewidth{700pt}
\tabletypesize{\scriptsize}
\tablehead{ Samples & 
\multicolumn{2}{c}{Star} & 
\multicolumn{2}{c}{Previous Work $^a$}  &
\multicolumn{2}{c}{Original Range65}  &
\multicolumn{2}{c}{Corrected Range15}   \\
& \colhead{TIC ID}  & \colhead{KIC ID}  
& \colhead{$\alpha_{\rm cum}$}  & \colhead{$\beta$}
&\colhead{$\alpha_{\rm cum}$}  & \colhead{$\beta$} 
&\colhead{$\alpha_{\rm cum}$}  & \colhead{$\beta$}
} 
\startdata
 & 102032397 & - & - & - & -1.436$\pm0.258$ & 48.52$\pm8.92$  & -0.938$\pm0.076$ & 31.51$\pm2.62$   \\
Sample-2 & 11046349 & - & - & - & -0.727$\pm0.072$ & 23.82$\pm2.48$  & -0.853$\pm0.051$ & 28.29$\pm1.75$   \\
&  ... & ... & ... & ... & ... & ... & ...& ...  \\
\hline 
& 150359500 & - & -1.144 & 36.766 & -1.036$\pm$0.029 & 33.00$\pm$0.93  & -1.018$\pm$0.014 & 32.70$\pm$0.44   \\
Sample-3 & - & 7131515 & - & - & -1.031$\pm$0.044 & 33.84$\pm$1.50  & -1.299$\pm$0.030 & 43.32$\pm$1.03   \\
& ... & ... & ... & ... & ... & ... & ...& ... \\
\enddata
\tablecomments{``Range65'' means the fitting within the energy range $P_{\rm det} \geq65\%$. `Range15'' means the fitting within the energy range $P_{\rm det} \geq15\%$. The flag ``-" indicates no available value. \\
$^{\rm a}$ From \citet{2020AJ....159...60G}\\
(This table is available in its entirety in machine-readable form.)}
\end{deluxetable*}
%\end{longrotatetable}

\section{Results}  \label{sect:results}
\subsection{Kepler Flaring Stars with TESS Light Curves}\label{sect:results_1} \label{subsect:results_twomission}
 For the Kepler flaring stars cross-matched with TESS Target List, we selected 13 targets (Sample-1) to conduct injection and recovery tests and correct every event using the detection probability and energy recovery ratio. 
 We perform a more refined source screening to compare  $\alpha_{\rm cum}$ obtained from Kepler and TESS flares, respectively.
 
Many fewer flare events are detected by the TESS light curve than by Kepler due to precision and the length of observation. Thus some stars only include a few flares with limited energy ranges. TIC 27685142 (KIC 11872364) does not have enough flare events to fit a cumulative FFD. Therefore, we excluded it from the subsequent analysis.

To avoid contamination by blended flaring stars, we also exclude stars with nearby sources. For the Kepler mission, about 10$\%$ of flares are probably polluted by nearby stars because some sources are very close to each other on CCDs (\citealp{2013ApJS..209....5S, 2016ApJS..224...37G}). 
TESS has a larger pixel scale and causes more contamination by blended flaring stars.
We cross-matched the \citet{2019ApJS..241...29Y}’s catalog with the TESS Target List using a distance of 21'' (the TESS pixel scale) to check for contamination by nearby stars. In these 13 targets we selected, we found that three have nearby stars within 21'', i.e. TIC 120629872 (KIC 2692704), TIC 406952118 (KIC 12646841) and TIC 120637232 (KIC 2692708).
In the analysis of cumulative FFDs, we also exclude these three stars. Thus, there are six stars left for further analysis. To extend the energy range of FFDs, we use the whole energy range to fit $\alpha_{cum, TESS}$ in both original and corrected cases, while the energy range to fit $\alpha_{\rm cum, Kepler}$ is chosen as $P_{\rm det}>15\%$.

\begin{figure}[ht!]
%\plotone{TIC27454242doubleMission_0828.pdf}
\centering
\includegraphics[width=0.8\textwidth]{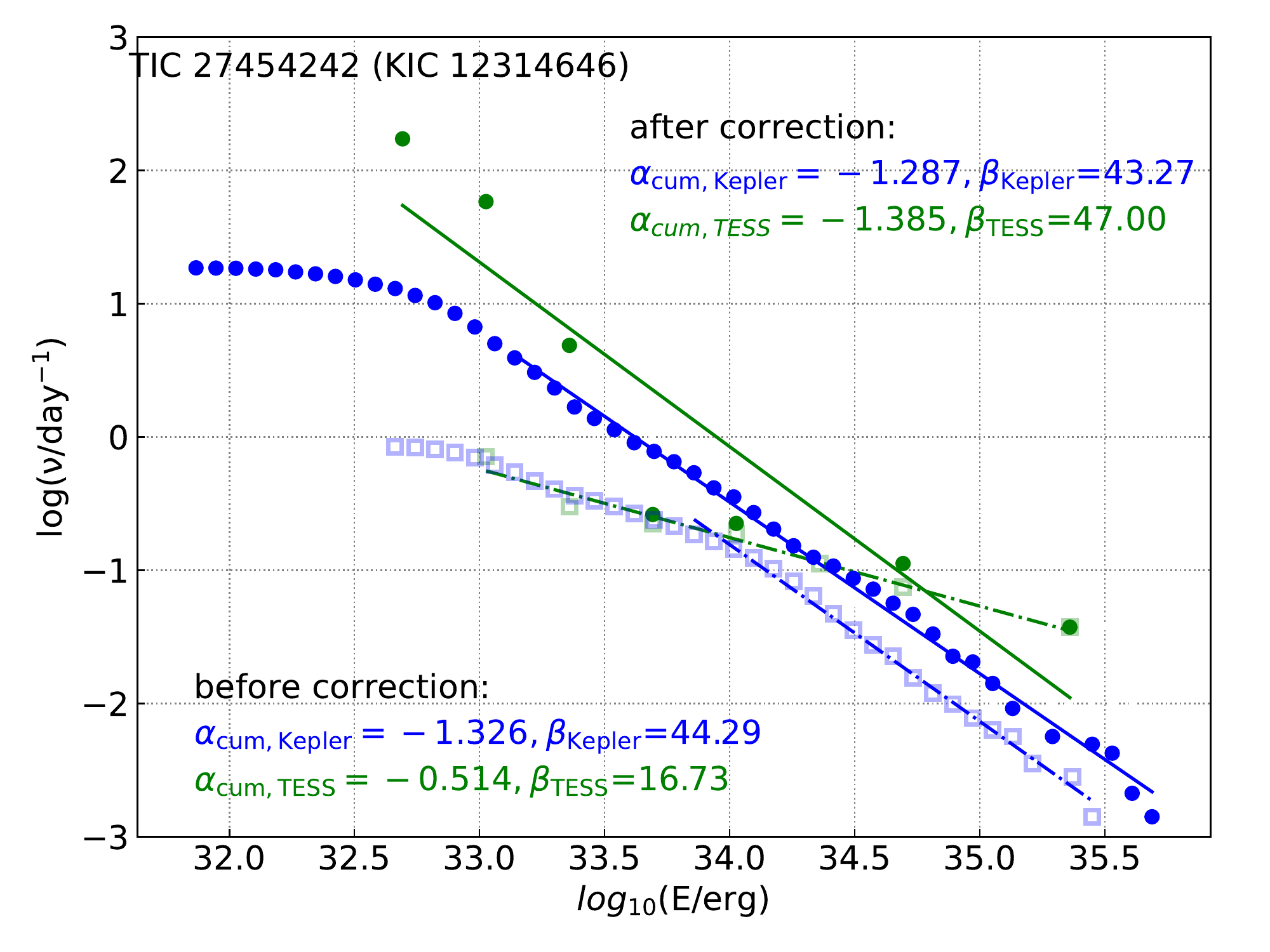}
\caption{The cumulative FFD of TIC 27454242 (KIC 12314646) before and after correction. The open green square is the original cumulative FFD detected from TESS data by our pipeline, and the green dot is the cumulative FFD after correction for each flare event using the detection probability and energy recovery ratio. The dashed-dotted and solid green lines indicate the fitted lines before and after correction, respectively. Similarly, for flare events detected from Kepler data, we use blue to present FFDs before and after correction.
\label{fig:single_twomission}}
\end{figure}

As a typical case, the cumulative FFDs of TIC 7454242 (KIC 12314646) from the TESS and Kepler mission are shown in Figure \ref{fig:single_twomission}. The corrected FFDs are much closer than the original FFDs, which indicates that the corrections adopted in this paper are useful for correcting the detecting bias from Kepler and TESS data.  

\begin{figure}[ht!]
%\plotone{diffMean.pdf}
\centering
\includegraphics[width=0.7\textwidth]{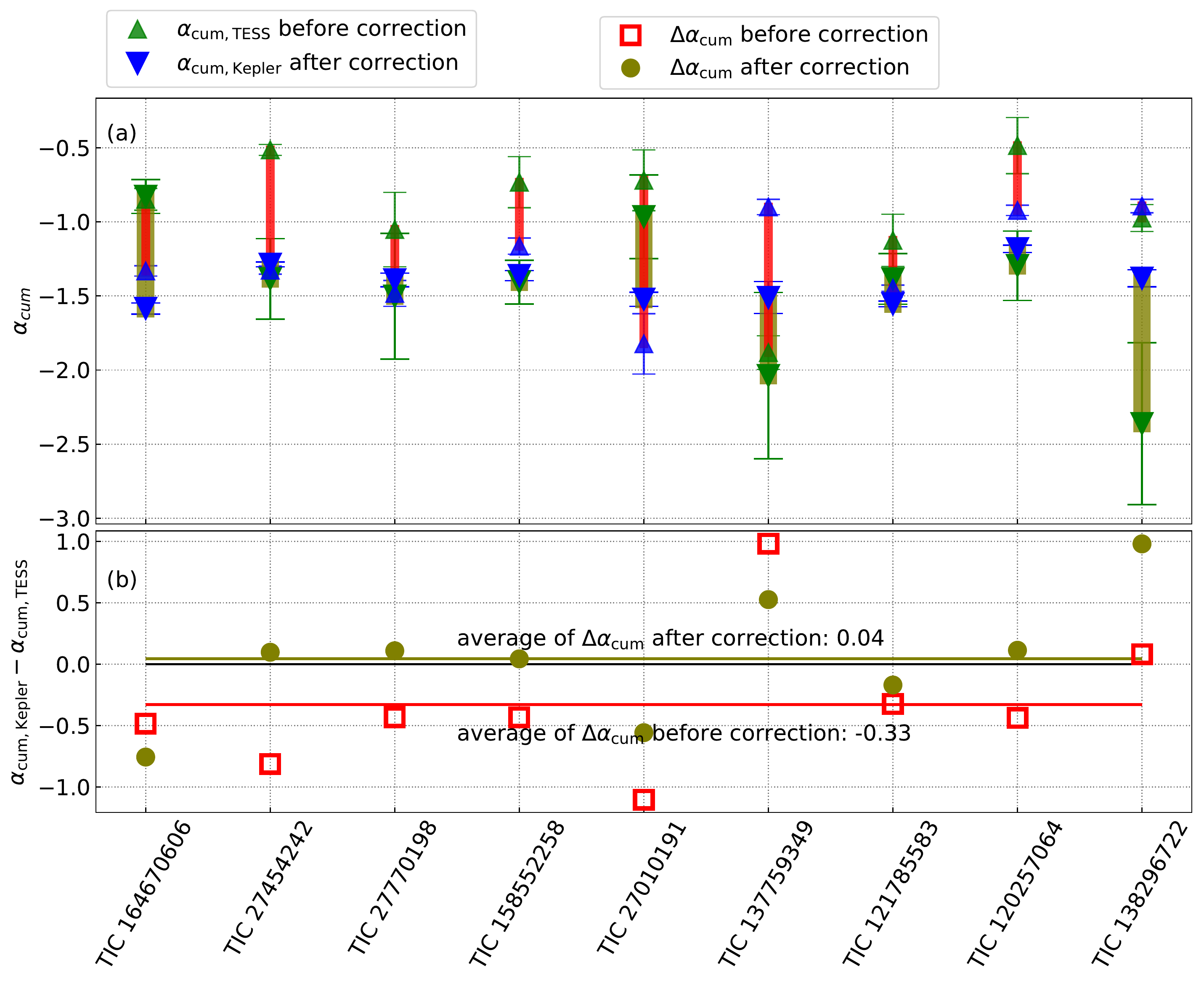}
\caption{ (a) Comparison of $\alpha_{\rm cum}$ values obtained by TESS and Kepler missions before and after correction. The regular and inverted green triangles indicate the $\alpha_{\rm cum}$ values obtained by fitting the cumulative FFDs from the TESS mission, before and after correction, respectively. Likewise, the blue ones come from the Kepler Mission. 
(b) The difference between the $\alpha_{\rm cum}$ values from the two missions ($\Delta \alpha_{\rm cum}=\alpha_{\rm cum, Kepler}-\alpha_{\rm cum, TESS}$), before and after correction. The open olive squares and solid red dots indicate differences of six stars, before and after correction, respectively. The red and olive horizontal lines indicate the average values of the difference before and after correction, which are -0.33 and 0.04, respectively. 
Note that because the number of flare events detected by the TESS light curve is small, we fit the cumulative FFD in the whole energy range for Sample-1 stars obtained from the TESS light curve, while the original $\alpha_{\rm cum, Kepler}$ come from the fitting of the original cumulative FFD with the energy range of $P_{\rm det} \geq 65\%$, and the corrected $\alpha_{\rm cum, Kepler}$ come from the fitting of the corrected cumulative FFD with the energy range of $P_{\rm det} \geq 15\%$.
\label{fig:difftwomiss}}
\end{figure}

For the remaining \textbf{nine} stars, Figure \ref{fig:difftwomiss} shows a comparison of the $\alpha_{\rm cum}$ values of TESS and Kepler missions obtained before and after correction. Nearly all original $\alpha_{\rm cum}$ obtained from TESS data are larger than those from Kepler, which indicates the obvious incomplete detection of flares with low energies via TESS. Additionally, TESS tends to detect flares with larger energy due to its redder observational band than Kepler's (\citealp{2019MNRAS.489..437D, 2021ApJS..253...35T}). Before correcting for flare events, we get an average difference of -0.33 between the two missions ($\Delta \alpha_{\rm cum}=\alpha_{\rm cum, Kepler}$-$\alpha_{\rm cum, TESS}$), while this decreases to 0.04 after correction. This indicates that the slope of cumulative FFDs obtained from TESS and Kepler missions is consistent.

After the correction, the $\alpha_{\rm cum}$ of some targets obtained by TESS and Kepler data are still different, e.g. TIC 27010191, TIC 137759349 and TIC 138296722. We speculate that the FFDs have changed during different observational epochs from Kepler and TESS.

\subsection{Solar-type Stars}\label{sect:results_2}
\begin{figure}[ht!]
\plotone{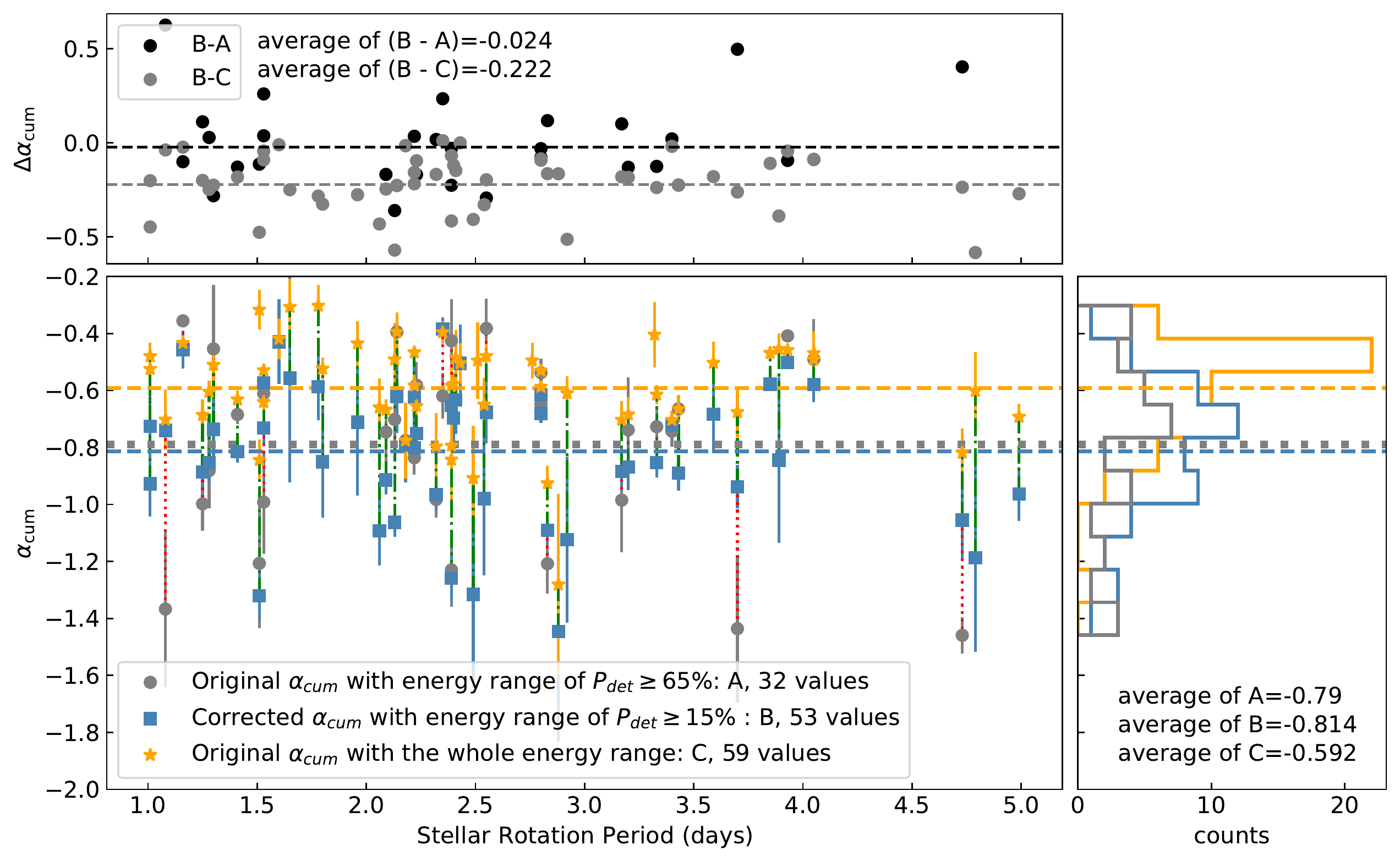}
\caption{Main panel: The $\alpha_{\rm cum}$ of Sample-2 stars before and after correction with different energy range fitting. 
The gray dots come from the original cumulative FFDs with an energy range of $P_{\rm det}\geq65\%$, labeled as A; their average, -0.79, is indicated by a gray horizontal dotted line.
The blue squares come from the corrected cumulative FFDs fitted with an energy range $P_{\rm det}\geq15\%$, labeled as B; their average, -0.814, is indicated by a blue horizontal dashed line.
The orange pentagrams come from the original cumulative FFDs fitted with the whole energy range, labeled as C; their average , -0.592, is indicated by an orange horizontal dashed line.
Right panel: the histogram of $\alpha_{\rm cum}$. The horizontal lines and colors are the same as in the main panel.
Top panel: the difference in $\alpha_{\rm cum}$ obtained from different data sets. 
The black dots are (B - A), and the gray dots are (B - C).
The averages of (B-A) and (B-C) are -0.024 and -0.222, and marked using black and gray horizontal dashed lines, respectively.
\label{fig:solartypealp}}
\end{figure}

For the 84 solar-type flaring stars from \citet{2020ApJ...890...46T}'s catalog (see Section \ref{subsect:sample_diffstars}), 32, 53, and 59 stars have enough flare events to fit cumulative FFDs individually when the fitting energy range is set to $P_{\rm det}\geq65\%$, $\geq15\%$ and the whole energy range, respectively. We fit these $\alpha_{\rm cum}$ in three different energy ranges, as shown in Figure \ref{fig:solartypealp}, and denote the three setting groups with different $P_{\rm det}$ as A, B, and C, respectively.

The average $\alpha_{\rm cum}$ in group A and B are nearly the same with a tiny difference of 0.024 (see the top panel of Figure \ref{fig:solartypealp}), which indicates that the completeness is improved in the energy range $P_{\rm det}>15\%$, via the correction method in this paper.
The average $\alpha_{\rm cum}$ in group C is $\sim0.2$ larger than that in B or A, which means the flares with lower energy are still incomplete and only partially corrected by our method. 

Based on the updated stellar parameters from the GAIA DR3 catalog, we removed the stars with effective temperatures not in the range of 5100$-$6000K, and possible binaries (RUWE$\textgreater$1.4) in the analysis of solar-type stars. We collect the flare events on \textbf{the remaining 53} stars to obtain a typical $\alpha_{\rm cum}$ of a solar-type star. $\alpha_{\rm cum}$ before and after correction is -1.24$\pm$0.10 and -1.08$\pm$0.04($P_{\rm det}>15\%$), respectively.

In previous works, both \citet{2019ApJS..241...29Y} and \citet{2020ApJ...890...46T} fit the FFD in the energy range above 10$34$ erg, although \citet{2020ApJ...890...46T} only focus on superflares. In this paper, the original detected number of flares peaks at $\sim10^{33.5}$ erg. The lower energy limit with $P_{\rm det}=15\%$ is extended to 10$^{33.0}$ erg. Thus our results of $\alpha_{\rm cum}$ are available in lower energy ranges. 

To compare with the result for the Sun and other previous works, we also fit the $\alpha$ of FFDs of all flare events on \textbf{the remaining 53} stars in Sample-2. I.e. $\alpha=-1.93\pm0.12$ with an energy range $P_{\rm det}\geq65\%$ before correction, and \textbf{$\alpha=-2.00\pm0.06$} with an energy range $P_{\rm det}\geq15\%$ after correction. 

After correction, $\alpha+1$ (-2.00+1) is closer to $\alpha_{\rm cum}$ (-1.08), which indicates that the FFD obeys the power-law distribution in the energy range $>10^{33.25}$ erg (see Section \ref{subsect:method_fitcffd}).
Our results are similar to previous works. For example, \citet{2021ApJ...906...72O} reported the statistical analyses of superflares on solar-type stars using all of the Kepler primary mission data and the updated Gaia DR2 catalog (\citealp{2018ApJ...866...99B}). After correction for gyrochronology and detection incompleteness, they obtained $\alpha=-1.5\pm0.1$ and $-1.8\pm0.1$ for fast rotating solar-type stars ($P_{\rm rot} \textless 5 $ days) with $T_{\rm eff}=5100-5600K$ and 5600$-$6000K, respectively, using the flare events with $10^{33.75}\ erg \ \textless\  E_{\rm bol}\ \textless 10^{35.75}\ erg$. \citet{2015EP&S...67...59M} and \citet{2013ApJS..209....5S} find that $\alpha$ of superflares with energy 10$^{33-36}$ erg and 10$^{34-35}$ erg on solar-type stars from Kepler vary from -1.5 and -2.2, respectively, since they consider a wider range of rotation period without correction of detection bias. Based on the first- and second-year observation of TESS, \citealp{2020ApJ...890...46T} and \citealp{2021ApJS..253...35T} obtained  $\alpha=-2.16\pm0.10$ (10$^{34.5-37.0}$ erg) and $\alpha=-1.76\pm0.11$ (10$^{34-36}$ erg), respectively.

%\textbf{Flares with different energy ranges is different. In previous study of Sun,\citet{2000ApJ...535.1047A} obtained $\alpha=-1.79$ of solar nanoflares (10$^{24-26}$ erg). \citet{1993SoPh..143..275C} got $\alpha=-1.53$ due to solar microflares (10$^{27-32}$ erg), while  \citet{1995PASJ...47..251S} gave an  $\alpha=-1.74$ for solar microflares with energy of 10$^{27-29}$ erg. Thus, the previous works and this paper combine all flares with different energies on different stars to fit the $\alpha$.}

%For other solar type stars, previous works show similar $\alpha$ values, and
%indicate the same physical mechanism for flares on solar type stars (\citealp{2019ApJ...880..105A,2021ApJS..253...35T}). 
%remove the reference!??

%Based on the first and second-year Observation of TESS, \citealp{2020ApJ...890...46T} and \citealp{2021ApJS..253...35T} obtained  $\alpha=-2.16\pm0.10$ (10$^{34.5-37.0}$ erg) and $\alpha=-1.76\pm0.11$ (10$^{34-36}$ erg), respectively. To sum up, previous works of both Sun and solar-type stars show that flares in different energy range result in different $\alpha$, i.e. the observed FFD maybe not the same power-law distribution exactly.

Fitting $\alpha$ due to all flares collected on a group of stars assumes different weights for each star based on flare counts. It may be dominated by stars with many more flares. Even if all stars have the same $\alpha$, the stacked FFD for all flares on these stars can have a different slope， because of the different counts and detectable energy range of FFDs for different stars. Thus, we also calculate the averaged $\alpha=-1.55\pm0.29$ ($\alpha_{cum}=-0.84\pm0.24$), assuming  equal weights for each star. The results is larger than the fitted value when combining all flares, but is consistent within 2 $\sigma$ uncertainty.

\subsection{Stars of Different Spectral Type } \label{sect:results_3}
Unlike solar-type stars, stars with lower temperatures, especially M dwarfs, are usually more active. In this section, according to the updated stellar parameters from GAIA DR3, we collect \textbf{22} stars in Sample-3 and eight stars in Sample-1 (Section \ref{subsect:results_twomission}), as well as 45 stars in Sample-2. Note that the total of 75 flaring stars from TESS and Kepler have stellar rotating period in the range 1 to 5 days, with only two exceptions-- TIC 121785583 (KIC 4355503) and TIC 120257064 (KIC 3934090) from Sample-1 (with the stellar rotation periods of 5.45 and 5.82 days). 

\begin{figure}[ht!]
\plotone{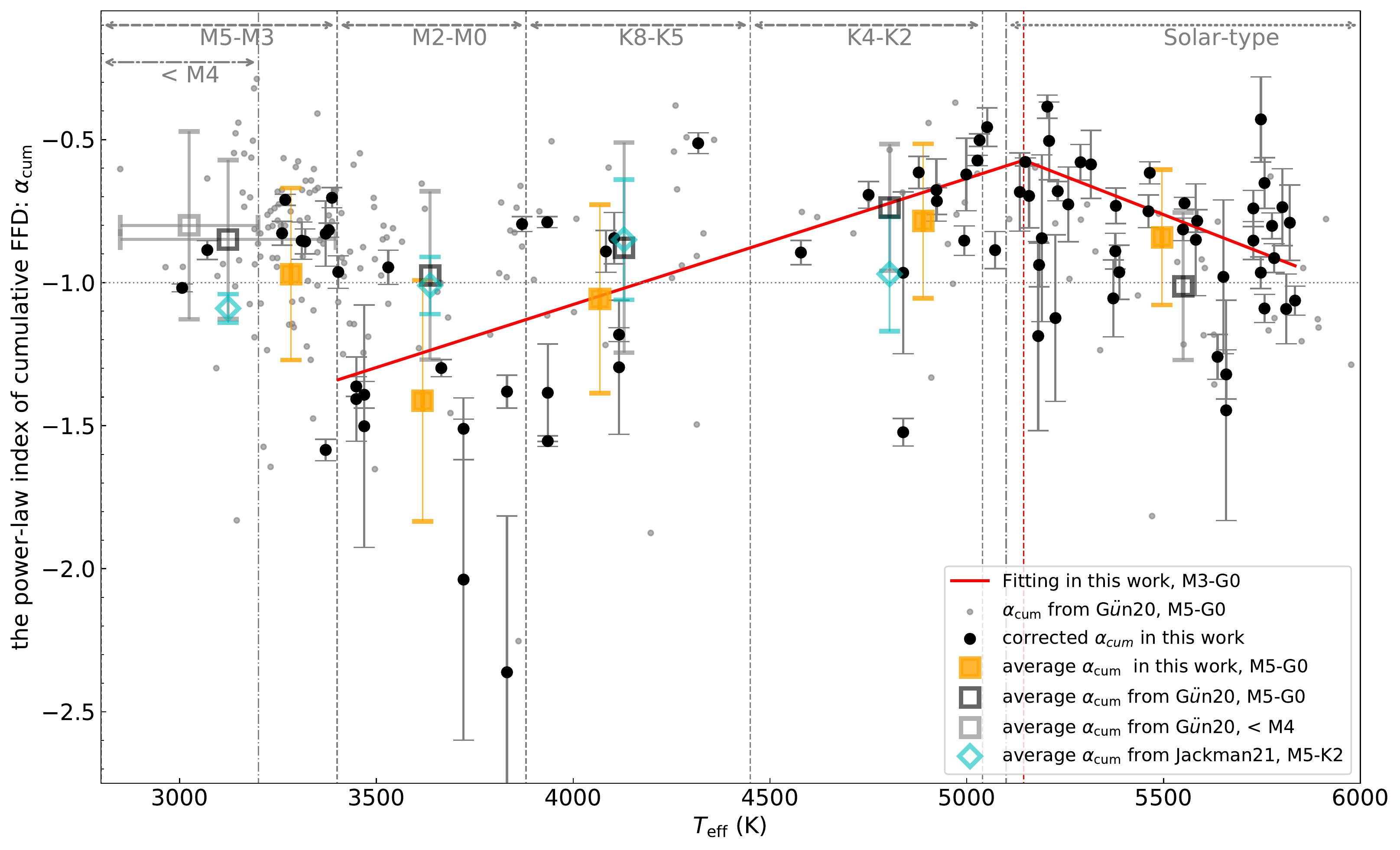}
\caption{$\alpha_{\rm cum}$ of stars of different subsets.
The gray dots are the values of  $\alpha_{\rm cum}$ from \citet{2020AJ....159...60G}.
The black dots mark those obtained by the corrected cumulative FFD fitting from the TESS and Kepler missions;
vertical dashed lines mark the different spectral type classifications.
The orange square marks the average $\alpha_{\rm cum}$ of each spectral type subset in this work.
The open black square marks the average $\alpha_{\rm cum}$ of \citet{2020AJ....159...60G} in each spectral type subset. The y-axis error marks the standard deviation of $\alpha_{\rm cum}$ in this subset. The average $\alpha_{\rm cum}$ of the spectral types that are cooler than M4 is marked by open gray square in particular.
The open cyan diamond indicates the values from \citet{2021MNRAS.504.3246J}. The solid red line marks the piecewise linear formulae (Eqn. \ref{equation:alpvsteff}) fitting the distribution relationship of $\alpha_{\rm cum}$ with $T_{\rm eff}$ from M2 to G1 stars. The Pearson correlation between $\alpha_{\rm cum}$ and $T_{\rm eff}$ is R=0.63 and R=-0.27 for stars from M2 to K1 and from K0 to G1, respectively. The coefficient of determination for fitting linear formulae are 0.39 and 0.10 in these two segments, respectively.}
\label{fig:alpvsteff}
\end{figure}

To compare with different previous works,
we calculate the average value of $\alpha_{\rm cum}$ from \citet{2020AJ....159...60G}’s catalog for each stellar type  and take the standard deviation of each type as the uncertainty. 
\citet{2021MNRAS.504.3246J} gives the FFD fits of main-sequence stars with the Next Generation Transit Survey (NGTS, \citealp{2018MNRAS.475.4476W}), providing a positive correlation between $\alpha_{\rm cum}$ and the stellar types later than K5.   
The values from previous works are listed in Table \ref{tab:alps_previous} and shown in Figure \ref{fig:alpvsteff}. 

\begin{deluxetable*}{ccccccc}
\tablecaption{Average $\alpha_{\rm cum}$ of Stars of Different Spectral Type Obtained from Previous Work and This Work.}
%\tablewidth{700pt
\tabletypesize{\scriptsize}
\tablehead{ Spectral type & Intermediate $T_{\rm eff}$  (K)& 
\citet{2020AJ....159...60G} $^{\rm a}$ & 
\citet{2021MNRAS.504.3246J} $^{\rm b}$ & 
This Work $^{\rm a}$ &
From Equation\ref{equation:alpvsteff} &
$N_{\rm star}$ in This Work
}
\startdata
Solar-type & 5550.0 & -1.01$\pm$0.26 & - & -0.84$\pm$0.24 & -0.79$\pm$0.26 & 38 \\
K4-K2 & 4804.0 & -0.74$\pm$0.22 & -0.97$\pm$0.20 & -0.78$\pm$0.27 & -0.72$\pm$0.08 & 10 \\
K8-K5 & 4129.0  & -0.88$\pm$0.37 & -0.85$\pm$0.21 & -1.06$\pm$0.33 & -1.02$\pm$0.07 & 6\\
M2-M0 & 3636.5 & -0.97$\pm$0.29 & -1.01$\pm$0.10 & -1.41$\pm$0.42 & -1.24$\pm$0.07 & 8\\
M5-M3 & 3123.0 & -0.85$\pm$0.28 & -1.09$\pm$0.05 & -0.97$\pm$0.30 & - & 10\\
Cooler than M4 & - & -0.80$\pm$0.33 & - & -0.95$\pm$0.07 & - & 2
\enddata
\tablecomments{$N_{\rm star}$ means the number of flaring stars; ``-" indicates no available value.\\
$^{\rm a}$ We calculate the average $\alpha_{\rm cum}$ in each subset and take the standard deviation as its error.\\
$^{\rm b}$ \citet{2021MNRAS.504.3246J} fitted an average  $\alpha$ combining flares on all stars, and adopted $\alpha_{\rm cum}$ value as $\alpha+1$, based on the assumption that all stars taken together were equivalent to a single, average, star (see the reference for detail).}
\label{tab:alps_previous}
\end{deluxetable*}
%\end{longrotatetable}

The average $\alpha_{\rm cum}$ of the targets from \citet{2020AJ....159...60G} based on TESS is independent of spectral type with Pearson correlation $R=-0.074$. Their average $\alpha_{\rm cum}$ on K8-K5 or M2-M0 stars is consistent with those from \citet{2021MNRAS.504.3246J}. Note that both previous works do not correct the flare events via injection and recovery. After correction, our work provides $\alpha_{\rm cum}$ with a more obvious correlation to the stellar type for stars cooler than K1. The Pearson correlation between $\alpha_{\rm cum}$ and $T_{\rm eff}$ is R=0.63, if we fit the positive trend as shown in Figure \ref{fig:alpvsteff}.
 
For solar-type stars, there is also a negative trend of $\alpha_{\rm cum}$, which is also revealed in Table 4 of \citet{2021ApJ...906...72O}) for fast rotating solar-type stars. Therefore, we use a piecewise linear formula to fit the trends of $\alpha_{\rm cum}$, in the range 3400-5900 K. The fitted linear formula is as follows:

\begin{equation}\label{equation:alpvsteff}
\alpha_{cum}= \left\{ \begin{array}{rcl}
0.44(\pm0.01)*\frac{T_{\rm eff}}{1000 \ K}-2.84(\pm0.05) & \mbox{for}
& 3400\ K\leq T_{\rm eff} < 5150\ K \\ 
-0.53(\pm0.03)*\frac{T_{\rm eff}}{1000\ K}+2.18(\pm0.18) & \mbox{for} & 5150\ K\leq T_{\rm eff}\leq 5900\ K
\end{array}\right.
\end{equation}
\\
{Note that the coefficients of determination are 0.39 and 0.10 in the different segments, respectively. Thus, more data from solar-types stars with well-determined parameters are crucial for validating the negative trend of $\alpha_{\rm cum}$ for solar-type stars.

According to the fitted correlation, $\alpha_{\rm cum}$ increases from -1.34 for an M2 star ($T_{\rm eff}$=3400K) to -0.57 for a K1 star ($T_{\rm eff}$=5150 K). This predicts a higher frequency of lower-energy flares in cooler stars with thicker convective envelopes. 
Our result is consistent with the $\alpha-\Omega$ dynamo theory, in which the magnetic field depends on the depth of the convective envelope (e.g., \citealp{1978mfge.book.....M, 1979cmft.book.....P}).
Conversely, for stars hotter than K1 ($T_{\rm eff}$ $\sim$5150 K), $\alpha_{\rm cum}$ starts to decrease with $T_{\rm eff}$ instead. For a solar-type star with $T_{\rm eff}= 5550K$, Equation \ref{equation:alpvsteff} provides $\alpha_{\rm cum}=-0.79$, which is consistent with the average value of $\alpha_{\rm cum}=-0.84\pm0.24$ in Sample-2. For stars with $T_{\rm eff}$ in the range between 4170 and 5940 K, the calculated $\alpha_{\rm cum}>$-1.0 according to Eqn. \ref{equation:alpvsteff}, thus the total energy is mostly contributed by flares with higher energy.

%If we fit $\alpha_{\rm cum}$ via the cumulative FFD of all flares on solar-type stars, \textbf{$\alpha_{\rm cum}$ are -1.24$\pm$0.10 and -1.08$\pm$0.04,} before and after correction, respectively. The inconsistence can be interpreted \textbf{according to the discussion in Section \ref{sect:results_2}}. We prefer the average $\alpha_{\rm cum}$ as the natural index of solar type stars in Sample-2.

According to the updated stellar parameters from GAIA DR3, there are only two fully convective stars (cooler than M4 type) in our samples, with average $\alpha_{\rm cum}$=-0.95. Eight M3 stars in our sample also show a similar $\alpha_{\rm cum}$ with an average value of -0.90.  There are 26 fully convective stars with a stellar period of 1.0-5.0 days in \citet{2020AJ....159...60G}’s catalog, and the average $\alpha_{\rm cum}$=-0.80 is represented by the open gray square to the left of the vertical dashed-dotted line in Figure \ref{fig:alpvsteff}. It is a little larger than the average $\alpha_{\rm cum}$ of M0-M2 type stars from both \citet{2020AJ....159...60G}, \citet{2021MNRAS.504.3246J} and this work. 
This indicates that the flare burst mechanism of fully convective stars may differ from other stars. The geometry of their magnetic fields is different and may have a switch in the dynamo (e.g., \citealp{2011ASPC..448..505S,2017NatAs...1E.184S}). That is why we exclude the 10 stars cooler than M2 in our sample to obtain the  Equation \ref{equation:alpvsteff}.

\section{Summary and Discussion}  \label{sect:summary}
Based on Kepler and TESS data, we consider the flare detection bias due to different missions and focus on FFD correction for flaring stars. We choose three samples with stellar rotation period ranging from 1 to $\sim$5 days in Section \ref{sect:samples} according to the flaring catalogs from \citet{2020AJ....159...60G}, \citet{2020ApJ...890...46T}, and \citet{2019ApJS..241...29Y}.

We built an uniform process to detect flares and to improve the parameters of flare events in Section \ref{sect:method}. 
Specifically, we select a suitable window width related to the stellar rotation period and detrend the light curve before finding flares in Section \ref{subsect:method_detrendlc}. Parametric thresholds are adopted due to photometric precision, and we calculate the flare energy by assuming a blackbody spectrum, and compare the completeness with previous works in Section \ref{subsect:method_findflares}. Each flare events are corrected by the detection probability $P_{\rm det}$ and energy recovery ratio $R_{\rm rec}$ in Section \ref{subsect:method_correction},  which are obtained from an injection and recovery test in Section \ref{subsect:method_inject}. Finally, we can fit the power-law index via the corrected cumulative FFD in Section \ref{subsect:method_fitcffd}, down to a lower energy limit with $P_{\rm det}=15\%$. 

In Section \ref{sect:results_1}, we analyze 13 Kepler flaring stars with TESS light curves in Sample-1, and select nine stars with reliable FFDs to fit $\alpha_{\rm cum}$ from TESS and Kepler data, respectively. Table \ref{tab:alps_s1} and Figure \ref{fig:difftwomiss} show that original $\alpha_{\rm cum}$ from TESS are larger than those from Kepler because of the less complete detection of low-energy flares for TESS.
The corrected $\alpha_{\rm cum}$ for most stars from Kepler and TESS become closer, as shown in Figure \ref{fig:difftwomiss}. We conclude that the slope of FFD or cumulative FFD (i.e., $\alpha$ or $\alpha_{\rm cum}$) after correction reveals more realistic values, and flares from different missions can be analyzed via the correction method to remove the detection bias. 

We extend a large number of low-energy flares through the detection and correction for 84 solar-type flaring stars from TESS in Section \ref{sect:results_2}. Thus we can reach a low-energy limit of 10$^{33}$ erg for cumulative FFD fitting on solar-type stars. Since the fitted $\alpha_{\rm cum}$ in the energy range $P_{\rm det}\geq65\%, P_{\rm det}\geq15\%$ and the whole energy range are similar, we conclude that the flares are nearly complete in the energy range $P_{\rm det}\geq15\%$ as shown in Figure \ref{fig:solartypealp}. Collecting all flares on 53 solar-type stars updated with stellar parameters from GAIA DR3, we obtain $\alpha_{\rm cum}$  $\sim$ -1.24$\pm$0.10 and -1.08$\pm$0.04 before and after correction, respectively, while the average $\alpha_{\rm cum}$ of all stars in Sample-2 is $-0.84\pm0.24$. The later value is preferred in this paper \textbf{(see Section \ref{sect:results_2})}, thus low-energy flares contribute less to the total flare energy on fast rotating solar-type stars.
%\textbf{The $\alpha_{\rm cum}$ is 0.77 and 0.95 for fast rotating stars ($P_{\rm rot}$=1-5 days ) with $T_{\rm eff}=5100-5600$ K and $T_{\rm eff}=5600-6000$ K, respectively. And the trend between these two $T_{\rm eff}$ bins is consistent \citet{2021ApJ...906...72O}.}

In Section \ref{sect:results_3}, based on stars in three samples, we obtain the $\alpha_{\rm cum}$ of stars from M5 to G1 through the correction of cumulative FFDs. Compared with previous works, we found that the $\alpha_{\rm cum}$ positively correlates to the stellar effective temperature for stars between M2 and K1. For stars hotter than K1, $\alpha_{\rm cum}$ may start to decrease with $T_{\rm eff}$ slightly. A piecewise linear empirical formula is obtained to monitor the correlations as shown in Figure \ref{fig:alpvsteff} and Eqn. \ref{equation:alpvsteff}. Assuming the spectrum is the same for all flares, for stars hotter than K5 stars and cooler than G0 stars, the total energy is mostly contributed by flares with higher energy. Fully convective stars cooler than M3 show a larger $\alpha_{\rm cum}$ than other M stars, which might indicate different dynamo mechanisms.

$\alpha_{\rm cum}$ is an excellent index to describe the FFD of a star. Coincidentally, one star (TIC 158552258 or KIC 7350067) of our Sample-1 has a confirmed super-Earth: Kepler-1646 b (\citealp{2016ApJ...822...86M}). The flaring host star (M dwarf) needs to be investigated more carefully to study the influence of flares on the super-Earth, and thus help us to know the location of the ``abiogenesis zone'' proposed by \citealp{2018SciA....4.3302R,2020AJ....159...60G}. 

In this work, we only study the fast rotating stars, because the rotation period is cut to 1 to $\sim$5 days. The rotation period depends on stellar age, which is also a crucial factor in determining stellar activity. As \citet{2011ApJ...743...48W} and \citet{2019ApJS..241...29Y} etc. show, flare activity of stars has a broken power law at the transitional Rossby number around 0.13. The Rossby number is the ratio of the stellar rotation period to the convective turnover time.  Thus this is the reason why there is a large scatter of $\alpha_{\rm cum}$ for stars of different types. To obtain better control samples, flaring stars need to be selected more carefully due to various parameters, if a much larger population of flaring stars is obtained in the near future.   

Additionally, the flare energy is calculated via simple models, i.e., the assumption of black-body radiation or a flat spectrum. Since Kepler and TESS have similar observational bands, thus we can correct the difference in FFD very well. However, it does not always work, especially for missions in quite different bands. Previous studies only discuss spectra from several early/mid M-dwarf flare stars (e.g., \citealp{2016ApJ...820...95K,2018csss.confE..42K,2019ApJ...871..167K}). So far, we do not have enough knowledge of the spectra of G-dwarf flares. Time series spectral observation and better physical models of flares will benefit us in understanding the precise flare energy distribution at different wavelengths. And we can combine the time series data to model the flare evolution more precisely, and achieve more accurate energy estimation for these flares.
%We published the code and the data for these analysis on github\footnote{\url{http://????}}
 
\begin{acknowledgments}
This work is supported by the National Natural Science Foundation of China (grant Nos. 11973028, 11933001, 1803012) and the National Key R\&D Program of China (2019YFA0706601). We also acknowledge the science research grants from the China Manned Space Project with NO. CMS-CSST-2021-B12 and CMS-CSST-2021-B09, as well as Civil Aerospace Technology Research Project (D050105).

\end{acknowledgments}


\vspace{5mm}
\facilities{TESS, Kepler}


\software{SciPy \citep{2020SciPy-NMeth}, AstroPy \citep{2013A&A...558A..33A,2018AJ....156..123A},  
          AltaiPony \citep{2016ApJ...829...23D,2021A&A...645A..42I}, 
          Lightkurve \citep{2018ascl.soft12013L}
          }



\appendix

\section{Parametric Criteria of Finding Flares}\label{append:flaredfind}
We detect flares following the parametric criteria in Equation (3a)-(3d) of \citet{2015ApJ...814...35C} as follows.

1. The flux of flare $f_i$, at epoch $i$, must be a positive excursion from the median quiescent flux value ($\bar {f_L}$). And the positive excess must be at least $N_{\rm 1}$ times larger than the local scattering of the light curve ($\sigma_L$), i.e.

\begin{equation}
\frac{f_i-\bar {f_L}}{\sigma_L} \geq N_1.
\end{equation}

2. The positive excess plus the photometric error $w_i$ at epoch $i$, must be greater than $N_{\rm 2}$$\sigma_L$, i.e. 

\begin{equation}
\frac{f_i-\bar {f_L} + w_i}{\sigma_L} > N_2  \label{equation:N2}.
\end{equation}

Approximately, the photometric error ($w_i$) and the local standard deviation ($\sigma_L$) can be considerated to be the same. The Equation \ref{equation:N2} can be written as

\begin{equation}
\frac{f_i-\bar {f_L}}{\sigma_L} +1 \geq  N_1 + 1  \label{equation:N2e}
\end{equation}

%The data points should satisfy the above criteria at the same time.
Therefore, when we set $N_2 \geq N_1 + 1$, the criterion on $N_1$ is satisfied automatically. 

3. To verify the confidence, the number of data points satisfying the above criteria must be at least $N_3$ continuously.

\section{Two methods to calculate flare energy}\label{append:energycalc}
There are two ways to calculate the flare energy. In this paper, we use the assumption of hot-blackbody radiation to calculate the flare energy (e.g., \citealp{2013ApJS..209....5S,2016ApJ...829...23D,2019ApJS..241...29Y,2020AJ....159...60G}, etc.). 

The luminosity of the star in the observing bandpass is
\begin{equation}
L_*=\pi R^2_* \int R_{\lambda}B_{\lambda}(T_{\rm eff})d\lambda,
\end{equation}

The luminosity of flare in the observing bandpass is

\begin{equation}
L_{\rm flare}(t)=A_{\rm flare}(t)\int R_{\lambda}B_{\lambda}(T_{\rm flare})d\lambda,
\end{equation}

where $R_{*}$ is the stellar radius and $R_\lambda$ is the response function in the observational band (TESS or Kepler band). $A_{\rm flare}(t)$ is the flare area while assuming that the flare temperature is constant throughout the duration. $B_{\lambda}(T_{\rm flare})$ and $B_{\lambda}(T_{\rm eff})$ are the Planck functions evaluated for the effective temperature of the flare and star, respectively.

The enhancement of the normalized light curve during flare events is
\begin{equation}
\frac{L_{\rm flare}(t)}{L_*}=\frac{A_{\rm flare}(t)}{\pi R^2_*}\frac{\int R_{\lambda}B_{\lambda}(T_{\rm flare})}{\int R_{\lambda}B_{\lambda}(T_{\rm eff})},
\end{equation}
which we can obtain from the observed light curves. We denote $\frac{\int R_{\lambda}B_{\lambda}(T_{\rm eff})}{\int R_{\lambda}B_{\lambda}(T_{\rm flare})}$ as $\eta$, which can be calculated through the response function of the observing bandpass and the effective temperatures of both star and flare. 

Finally, we arrive at the expression for the bolometric energy of the flare, given as

\begin{equation}
E_{flare}=\sigma_{\rm B} \cdot T^4_{\rm flare}\cdot \int A_{\rm flare}(t) dt=\sigma_{\rm B}\cdot T^4_{\rm flare} \cdot \pi R^2_* \cdot \eta \cdot \int \frac{L_{\rm flare}(t)}{L_*}dt
\end{equation}

where $\sigma_{\rm B}$ is the Stefan$-$Boltzmann constant. 
\\
\\
%\newlines

Another way to calculate flare energy is to integrate the energy based on the variations in the normalized light curve, assuming the variations in the observed band are the same as that in any other band (e.g., \citealp{2015ApJ...798...92W}). Thus, the bolometric energy of the flare is given as

\begin{equation}
E_{flare}=\sigma_{\rm B} \cdot 4\pi R^2_* \cdot T^4_{\rm eff}\cdot \int \frac{L_{\rm flare}(t)}{L_*}dt.
\end{equation}

Note that the calculated energy is usually larger than the blackbody assumption if we choose the flare temperature as 9000 K.

The calculated ratio of flare energies between the second method and the first is as follows:

\begin{equation}
r=\frac{4}{\eta}  \cdot (\frac{T_{\rm eff}}{T_{\rm flare}})^4
\label{equation:ratiooftwo}
\end{equation}

Taking a solar-like star ($T_{\rm eff}$=5800 K) and $T_{\rm flare}$=9000 K as examples, we estimate r = 2.93 in the TESS bandpass (600-1000 nm).


\bibliography{sample631}{}
\bibliographystyle{aasjournal}


}
\end{CJK*}
\end{document}